\documentclass[twocolumn]{article}
\usepackage{subfiles}
\usepackage{graphicx}
\usepackage{amsmath,amssymb,amsfonts}
\usepackage{algorithmic}
\usepackage{textcomp}
\usepackage{xcolor}
\usepackage{soul}
\usepackage{url}
\usepackage{authblk}
\usepackage{multirow}
\usepackage{array}
\usepackage[labelfont=bf]{caption}
\usepackage[capitalise]{cleveref}
\usepackage[labelformat=simple]{subcaption}

\usepackage[numbers,sort&compress]{natbib}

\begin{document}
\renewcommand{\thefootnote}{\fnsymbol{footnote}}
\title{How to Incorporate External Fields in Analog Ising Machines}
\def\correspondingauthor{\footnote{Correspondence and requests should be addressed to R.D.P (email: robbe.deprins@ugent.be) or to J.L. (email: Jacobus.Stijn.T.Lamers@vub.be).}}
\def\contributedequally{\footnote{These authors contributed equally.}}

\author[1,2]{Robbe De Prins \contributedequally{}\correspondingauthor{}}
\author[2]{Jacob Lamers $^*$$^\dagger$}
\author[1]{Peter Bienstman}
\author[2]{Guy Van der Sande}
\author[2]{Guy Verschaffelt}
\author[3]{Thomas Van Vaerenbergh}

\affil[1]{Photonics Research Group, Ghent University – imec, Technologiepark-Zwijnaarde 126, 9052 Gent, Belgium}
\affil[2]{Applied Physics Research Group, Vrije Universiteit Brussel, Pleinlaan 2, 1050 Brussels, Belgium}
\affil[3]{Hewlett Packard Labs, Diegem, Belgium}

\date{}
\maketitle
\renewcommand{\thefootnote}{\arabic{footnote}}

\begin{abstract}
Ising machines (IMs) are specialized devices designed to efficiently solve combinatorial optimization problems (COPs). They consist of artificial spins that evolve towards a low-energy configuration representing a problem's solution. Most realistic COPs require both spin-spin couplings and external fields. In IMs with analog spins, these interactions scale differently with the continuous spin amplitudes, leading to imbalances that affect performance. Various techniques have been proposed to mitigate this issue, but their performance has not been benchmarked.
We address this gap through a numerical analysis. We evaluate the time-to-solution of these methods across three distinct problem classes with up to 500 spins. 
Our results show that the most effective way to incorporate external fields is through an approach where the spin interactions are proportional to the spin signs, rather than their continuous amplitudes.
\end{abstract}

\section{Introduction}
\label{sec: intro}
Combinatorial optimization problems lie at the heart of numerous computational challenges in disciplines that range from industry to fundamental science. Some examples are problems in logistics \cite{SupplyChainLogisticsWithAnnealing}, finance \cite{Fincance},  biology \cite{Protein_Folding, ComputationalBiology}, job scheduling \cite{Job_Scheduling}, and traffic flow regulation \cite{Traffic_Flow}. Many of these problems are nondeterministic polynomial-time (NP) complete, making it exceptionally challenging to develop efficient algorithms to solve them. 
Today, such problems are demanding vast amounts of energy, with high-performance computing centers persistently dedicating substantial resources to them  \cite{gurobi_nfl_scheduling}.

One promising solution is the use of specialized hardware rather than general-purpose digital computers. 
More specifically, there has been a surge of interest in Ising machines (IMs): hardware implementations designed to find low-energy states of the Ising model, defined by the following Hamiltonian:
\begin{equation}
\mathcal{H}\left(\{\sigma_i\}\right)=-\frac{1}{2} \sum_{i,j}^N J_{ij} \sigma_i \sigma_j-\sum_i^N h_i \sigma_i.
\label{eq:ising hamiltonian binary}
\end{equation}
Here, $\sigma_i \in \{-1,1\}$ are Ising spins, and $N$ is the total number of spins. $J_{ij}$ and $h_i$ denote the real-valued strengths of the spin couplings and external fields, respectively.

It has been shown that it is possible to formulate any problem within the NP complexity class as an Ising model with only polynomial overhead \cite{karp2009reducibility, mezard1987spin}.
The ground state, i.e.~the configuration of binary spins with the lowest energy according to \cref{eq:ising hamiltonian binary}, then corresponds to the optimal solution of the optimization problem. 

Many IM implementations have been proposed so far \cite{OverviewMcMahon}. Some implementations are constructed using spins that are intrinsically \textit{binary}, such as implementations based on quantum annealing \cite{Quantum_annealing_DWave}, probabilistic p-bits \cite{Probabilistic_computing_with_pbits, Stochastic_pbits_for_invertible_logic}, spatial light modulators \cite{Conti}, and memristor Hopfield neural networks \cite{cai2020power}. Others utilize \textit{analog spins}, which take continuous values while satisfying constraints that enforce both the COP structure and spin bistability. Examples are Ising machines based on degenerate optical parametric oscillators \cite{2000nodes, 100000SpinsCIM}, opto-electric oscillators \cite{Poor_mans_CIM}, electrical resonators \cite{Ising_machine_based_on_networks_of_subharmonic_electrical_resonators}, memristor crossbar arrays \cite{jiang2023efficient}, and polariton condensates \cite{PolaritonCondensates, PolaritonCondensates_2}. In this paper, we focus on Ising machines with analog spins.

In general, analog IMs are modeled by a set of differential equations. The time evolution of the amplitude of spin $i$, denoted $s_i \in \mathbb{R}$,  is described as:
\begin{equation}
\frac{d s_i}{d t}=\mathcal{F}_i\left(s_i, \alpha, \beta I_i\right),
\label{eq:generic_update_eq}
\end{equation}
where $\mathcal{F}$ is a nonlinear function, $\alpha$ is the linear gain, $\beta$ is the interaction strength. $I_i$ is the local field of spin $i$, which can be modeled as follows:
\begin{equation}
    I_i = \sum_j J_{ij} s_j + h_i.
    \label{eq:original external fields}
\end{equation}
The IM’s energy, as given by \cref{eq:ising hamiltonian binary}, can be evaluated with spin amplitudes mapped to binary spins via $\sigma_i = \text{sgn}(s_i)$, where $\text{sgn}(\cdot)$ denotes the sign function.

While some studies have considered tasks that require external fields \cite{gunathilaka2023effective,ikeda2019application,zhang2022efficient,original_auxTrick_paper,original_meanAbsTrick_paper,MeanField_Annealing}, the emphasis in benchmarking IMs (both with binary and analog spins) remains largely on problems that do not require such fields, such as the Max-Cut problem \cite{commander2009maximum}. This focus contrasts with the requirements of most industrially relevant COPs, which necessitate external fields for their encoding into an IM \cite{Ising_formulations_of_many_NP_problems}.

For IMs with analog spins, implementing external fields can be challenging.
This is clear from \cref{eq:original external fields}, where as the spin amplitudes decrease, the spin coupling terms ($\propto J_{ij} s_j$) are weakened relative to the external fields ($\propto h_i$), potentially causing the latter to dominate. Conversely, when spin amplitudes exceed one, the couplings may dominate instead. Since embedding a COP into an IM requires careful tuning of the values of $J_{ij}$ and $h_i$, such imbalances may undermine the IM's performance.

In the past, various techniques have been proposed to mitigate this issue \cite{original_auxTrick_paper,original_meanAbsTrick_paper}, but their impact on the IM's performance has not yet been benchmarked. In this work, we address this gap by conducting a numerical study. 
Moreover, we include an alternative method that substitutes $J_{ij} s_j$ for $J_{ij}\,\text{sgn}(s_j)$ in \cref{eq:original external fields}.
This method was originally proposed for ballistic simulated bifurcation \cite{inspiration_idea_thomas,higher_order_dSB}, a specific type of analog IM where $\mathcal{F}$ in \cref{eq:generic_update_eq} is a \textit{linear} function (whereas a nonlinear function is used in this work) and spins evolve with momentum rather than following a gradient method. While this binarization method is uncommon in other analog IMs \cite{OverviewMcMahon} and has not been included in earlier studies of imbalances between spin couplings and external fields \cite{Mean-field_CIM_with_artificial_ZeemanTerms}, we will show that it is particularly effective for incorporating external fields and outperforms the previously proposed techniques.

We apply all methods to three different problem classes, each requiring a distinct mapping to embed problems into an Ising machine:  (a) Sherrington–Kirkpatrick (SK) Hamiltonians with random external fields, which we generated ourselves, (b) quadratic unconstrained binary optimization (QUBO) problems from BiqMac's Beasley benchmark set \cite{BiqMac}, and (c) Max-3-Cut problems, where graphs are generated using the Rudy generator \cite{rudy_graph_generator}. 
While SK problems serve as a general benchmark—sampling $J_{ij}$ and $h_i$ from Gaussian distributions rather than relying on a structured mapping—the other two classes represent typical ways external fields emerge in problem embeddings (see Results for details).

In addition to the techniques discussed above, we also explore a simple recalibration of the external fields by rescaling them with a constant factor, i.e.~replacing $h_i$ by $\zeta h_i$ in \cref{eq:ising hamiltonian binary,eq:original external fields} where $\zeta \in \mathbb{R}$. This adjustment essentially modifies the embedding process of COPs and can be easily combined with any of the prior methods. This constant rescaling has been proposed in previous works \cite{Mean-field_CIM_with_artificial_ZeemanTerms, original_meanAbsTrick_paper}, where it was suggested to enhance the performance of IMs.
In this work, we show that such a constant rescaling can indeed be necessary when encoding COPs with soft constraints, such as for Max-3-Cut problems. However, for COPs without such constraints, such as in the SK and Beasley problems, this rescaling removes the guarantee that the ground-state spin configuration solves the COP, resulting in an incorrect COP embedding.

\section{Modeling analog Ising machines}
\label{sec: Modeling_analog_IMs}
\subsection{Transfer function}
Ising machines with analog spins rely on nonlinear dynamics to establish bistable spin amplitudes. This bistability is typically realized through a pitchfork bifurcation, which can be achieved via different nonlinear functions $\mathcal{F}$ in \cref{eq:generic_update_eq}. A prior study compared many nonlinear functions \cite{Order_of_magnitude}, and showed that the following nonlinearity achieves the best performance:
\begin{equation}
    \frac{ds_i}{dt} = -s_i + \tanh\left(\alpha s_i + \beta I_i\right).
    \label{eq: sigmoid nonlin}
\end{equation}
This nonlinearity is commonly employed to incorporate the saturation of the spin amplitudes, often observed in experimental setups \cite{Poor_mans_CIM}. Its good performance can be attributed to its inherent suppression of amplitude inhomogeneity, a common source of error when mapping COPs to IMs with analog spins. 
Moreover, this nonlinearity is particularly well-suited when external fields are present. Indeed, as described in the introduction, external fields ($\propto h_i s_i$) dominate over the spin couplings ($\propto J_{ij} s_i s_j$) when spin amplitudes are small ($s_i \ll 1$). Conversely, spin couplings dominate when the spin amplitudes exceed one. \cref{eq: sigmoid nonlin} now constrains the spin amplitudes to the range $[-1, 1]$, thereby excluding the latter scenario.
For these reasons, the remainder of this work exclusively considers the hyperbolic tangent nonlinearity.

\subsection{Annealing scheme}
At the start of a simulation of the IM, the spin amplitudes are initialized near zero, and their evolution is tracked over time by numerically integrating \cref{eq: sigmoid nonlin}. Specific details are provided in Supplementary Note 1. The interaction strength $\beta$ is increased at each timestep, following a commonly used linear annealing scheme \cite{paper_Ganguli, PaperJacob_UsingContinuationMethods}:
\begin{equation}
    \beta(t) = \beta_0 + v_\beta t.
    \label{eq:LinearAnnealing}
\end{equation}
Here, $\beta_0$ is the initial $\beta$-value and $v_\beta$ is the annealing speed. As also detailed in Supplementary Note 1, we choose a relatively low and constant noise strength while integrating \cref{eq: sigmoid nonlin}.

\subsection{Methods to incorporate external fields}
\label{sec: definition tricks}
\begin{figure*}[tb]
    \centering
    \includegraphics[width=1.0\linewidth]{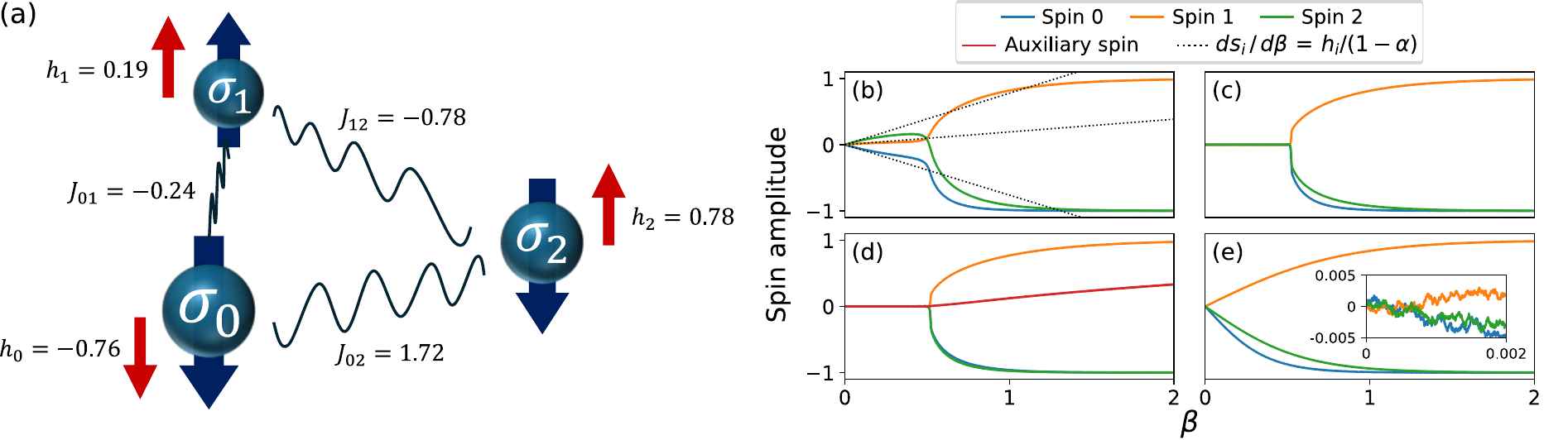}
    \caption{(a) Example of a 3-spin COP with ground state configuration (-1,1,-1). 
    (b)-(e) Spin evolution under the methods described in \cref{eq:original external fields,equation: aux trick,equation: meanAbs trick,eq:spin sign method}, respectively.
    In (b), the initial direction of the spins (dotted lines) is given by $h_i / (1-\alpha)$, independently of the spin-spin couplings $J_{ij}$. Methods in (c)-(e) are designed to prevent the external fields from dominating the spin-spin couplings. The inset in (e) shows a zoom-in near $\beta=0$.
    }
    \label{fig: conceptual: 3-spin SK}
\end{figure*}

In this manuscript, we will compare the performance of four methods for incorporating external fields, which we outline below. For each of these methods, we will illustrate the spin dynamics using a 3-spin COP, which is visualized in \cref{fig: conceptual: 3-spin SK}(a), and for which the ground state configuration is (-1,1,-1).

\subsubsection{The original external fields}
All considered methods essentially modify the local field of spin $i$, as originally defined by \cref{eq:original external fields}. Hence, we will further refer to \cref{eq:original external fields} as the `original external fields'.
For the 3-spin COP of \cref{fig: conceptual: 3-spin SK}(a), the evolution of the spin amplitudes under \cref{eq:original external fields,eq: sigmoid nonlin,eq:LinearAnnealing} is visualized in \cref{fig: conceptual: 3-spin SK}(b) where $\beta_0=0$. As we start to increase $\beta$, the spins initially follow the dotted lines. As shown in Supplementary Note 2, these dotted lines correspond to $\frac{d s_i}{d \beta} = \frac{h_i}{1-\alpha}$, such that the spin dynamics at low $\beta$ are determined by the external fields $h_i$, independent of the spin couplings $J_{ij}$. As the amplitudes grow, the spin couplings strengthen (the first term in \cref{eq:original external fields} grows), causing spin 2 to flip and leading to the ground state configuration (-1,1,-1). 

Thus, for this simple COP, the IM successfully overcomes the initial field-driven behavior. 
However, for more complex problems, this may not always be the case. As discussed in \cref{sec: intro}, an imbalance between the spin couplings and the external fields can undermine the validity of the COP embedding.

\subsubsection{The mean absolute spin method}
One way to address the imbalance between spin couplings and the external fields is to rescale the external fields with the mean of the absolute values of the spin amplitudes \cite{original_meanAbsTrick_paper}, leading to the following local field:
\begin{equation}
I_i = \sum_{j=1}^N J_{i j} s_j+h_i \frac{1}{N} \sum_k\left|s_k\right|.
\label{equation: meanAbs trick}
\end{equation}
This approach ensures that the external field terms scale linearly with the spin amplitudes, consistent with the scaling of the spin couplings. 

\cref{fig: conceptual: 3-spin SK}(c) shows the evolution of the spins following this approach. We observe that the spin amplitudes remain zero until they bifurcate into the ground-state solution. Hence, this method prevents the external fields from dominating over the spin couplings.

We observe that the origin ($s_i = 0$, $\forall i$) is a stable fixed point up to a certain $\beta$ value, at which the spins bifurcate. Ideally, $\beta_0$ would be set equal to this critical value. However, as explained in Supplementary Note 3, determining this exact value is challenging, and we further choose to set $\beta_0 = 0$ when using the mean absolute spin method. Although a heuristic could, in principle, estimate $\beta_0$ more accurately and potentially improve performance, we do not explore that direction here. As shown in Supplementary Note 3, the choice of $\beta_0$ has only a moderate effect and does not influence our conclusions in the remainder of this work.

\subsubsection{The auxiliary spin method}
An alternative method to rescale the external fields is provided by the `auxiliary spin method' \cite{original_auxTrick_paper}.
It introduces an auxiliary spin amplitude $s_{N+1}$, and replaces each external field $h_i$ by a coupling between spin $i$ and the auxiliary spin. The local field of spin $i$ is modified as follows:
\begin{equation} 
\begin{split} 
    I_i = & \begin{cases} 
    \sum_{j=1}^N J_{ij} s_j + h_i s_{N+1}, & \text{if } i < N+1, \\ 
    \sum_{j=1}^N h_j s_j, & \text{if } i = N+1, \end{cases} \\ 
    = & \sum_{j=1}^{N+1} J'_{ij} s_j, \quad \text{where } \mathbf{J'} =
    \begin{bmatrix}
    \mathbf{J} & \mathbf{h} \\
    \mathbf{h}^T & 0
    \end{bmatrix}~. 
\end{split} 
\label{equation: aux trick} 
\end{equation}
Hence, applying this technique only requires the implementation of spin couplings.

It can be shown that if $\mathbf{s}_\text{opt}$ is the optimal solution to the original Ising problem of \cref{eq:ising hamiltonian binary}, then $(\mathbf{s}_\text{opt}, 1)$ and $(-\mathbf{s}_\text{opt}, -1)$ are degenerate solutions of the system with the auxiliary spin \cite{original_auxTrick_paper}. Hence, if the auxiliary spin amplitude takes a negative value, all spins are flipped before evaluating the IM's energy via \cref{eq:ising hamiltonian binary}.

\cref{fig: conceptual: 3-spin SK}(d) illustrates the evolution of the spins using this method. Similar to the mean absolute spin approach, the spin amplitudes remain zero until they bifurcate into the ground-state solution. However, for the auxiliary spin method, we show in Supplementary Note 4 that the $\beta$-value at this bifurcation can be easily calculated, allowing us to set $\beta_0$ to this value.

\subsubsection{The spin sign method}
The spin sign method \cite{inspiration_idea_thomas,higher_order_dSB} defines the local field of spin $i$ as follows:
\begin{equation}
    I_i = \sum_j J_{ij} \, \text{sgn}(s_j) + h_i.
    \label{eq:spin sign method}
\end{equation}

\cref{fig: conceptual: 3-spin SK}(e) shows the evolution of the spins following this approach where we set $\beta_0=0$. The inset provides a more detailed view near $\beta_0$. For this simple COP, we observe that the spin sign method first explores the surroundings of the origin. Once the signs of the spins correspond to the COP's solution, the local fields remain at constant values such that the spin amplitudes continue to grow linearly until they saturate.

\subsection{Scaling external fields by a constant}
\label{sec:background:zeta-scaling}
Previous works have proposed to rescale the external fields by a constant $\zeta \in \mathbb{R}$ \cite{Mean-field_CIM_with_artificial_ZeemanTerms,original_meanAbsTrick_paper}, thereby substituting $h_i$ for $\zeta h_i$ in \cref{eq:original external fields,equation: meanAbs trick,equation: aux trick,eq:spin sign method}. However, these works did not provide any insights into why this rescaling technique should be adopted and for which COPs it is effective.

Embedding a COP into an IM determines the values of  $J_{ij}$ and $h_i$.
For many COPs, such as SK problems and Beasley problems, this embedding guarantees that the ground-state of \cref{eq:ising hamiltonian binary} corresponds to a solution of the COP. 
Hence, it is counterintuitive to set $\zeta \neq 1$, since this modifies the energy landscape of \cref{eq:ising hamiltonian binary}, thereby eliminating this guarantee. 
For example, if we substitute $h_i$ with $\zeta h_i$ in \cref{eq:ising hamiltonian binary} and set $\zeta \gg 1$, the external fields dominate the system, resulting in a ground-state configuration ${\text{sgn}(h_i)}$. In this scenario, the spin couplings -- which encode information that is essential to solving the COP -- are effectively disregarded.

For some problems, such as Max-3-cut, so-called soft constraints are used to embed a COP. These constraints are implemented by adding terms to the energy function of \cref{eq:ising hamiltonian binary}, such that violating these constraints incurs an energy penalty. These terms come with a prefactor that indicates the importance of the corresponding constraint. 
However, when using soft constraints, the ground-state configuration will only solve the COP if the prefactors of all constraints are set to appropriate values. Specifically, for Max-3-Cut problems, we will demonstrate that determining the correct prefactor is challenging for realistically-sized COPs, but surprisingly, this issue can be resolved by applying a factor $\zeta \approx 0.6$.

\section{Results}
\label{sec:results}
In this section, we discuss the performance of the rescaling techniques introduced in the previous section. We do this using three different problem classes, which are introduced in the following subsections.

The IM's performance is evaluated using the time-to-solution (TTS) metric, which represents the total time needed to reach the target state with 99\% probability. It is defined as:
\begin{equation}
    \label{eq:TTS}
    \text{TTS} = \begin{cases}
        T, & \text{if } P>0.99,\\
        T\frac{\log(0.01)}{\log(1-P)}, & \text{if } 0<P\leq0.99,\\
        \infty, & \text{if } P=0,
    \end{cases}
\end{equation}
where $P$ is the probability of reaching the target state, and $T$ is the (dimensionless) time window over which \cref{eq: sigmoid nonlin} is integrated. Whenever possible, the ground state is chosen as the target. When the ground state is unknown, a target energy is chosen corresponding to the best-known solution, as detailed in later sections.

Whereas $T$ is typically treated as a hyperparameter \cite{hamerly_comparison_Dwave_CIM}, which leads to the performance measure $\min_T \text{TTS}\left(T,P(T)\right)$, here we choose to set $T$ as the average simulation time required to reach the target state across successful runs. This choice provides a practical approximation of the TTS, as it demands fewer computational resources for evaluation, compared to treating $T$ as a hyperparameter. We estimate the values of $P$ and $T$ over 100 runs.

The performance of the IM depends
on the choice of the linear gain $\alpha$ and the annealing speed $v_\beta$ (cf. \cref{eq: sigmoid nonlin,eq:LinearAnnealing}). Given a COP and a method from \cref{sec: definition tricks}, these hyperparameters are optimized, and the corresponding value of  $\min_{\alpha,v_\beta}\text{TTS}$ is reported. For the Max-3-Cut problems, an additional hyperparameter specific to the Ising formulation is optimized (see later). 
Unless stated otherwise, the constant rescaling factor $\zeta$, introduced in \cref{sec:background:zeta-scaling}, is set to 1. Where relevant, it is treated as an additional hyperparameter. Further details about the hyperparameter optimization procedure are provided in the Supplementary Note 1. 

\subsection{Sherrington–Kirkpatrick Hamiltonians with random external fields}
\begin{figure*}[htb]
    \centering
    \includegraphics[width=1.0\linewidth]{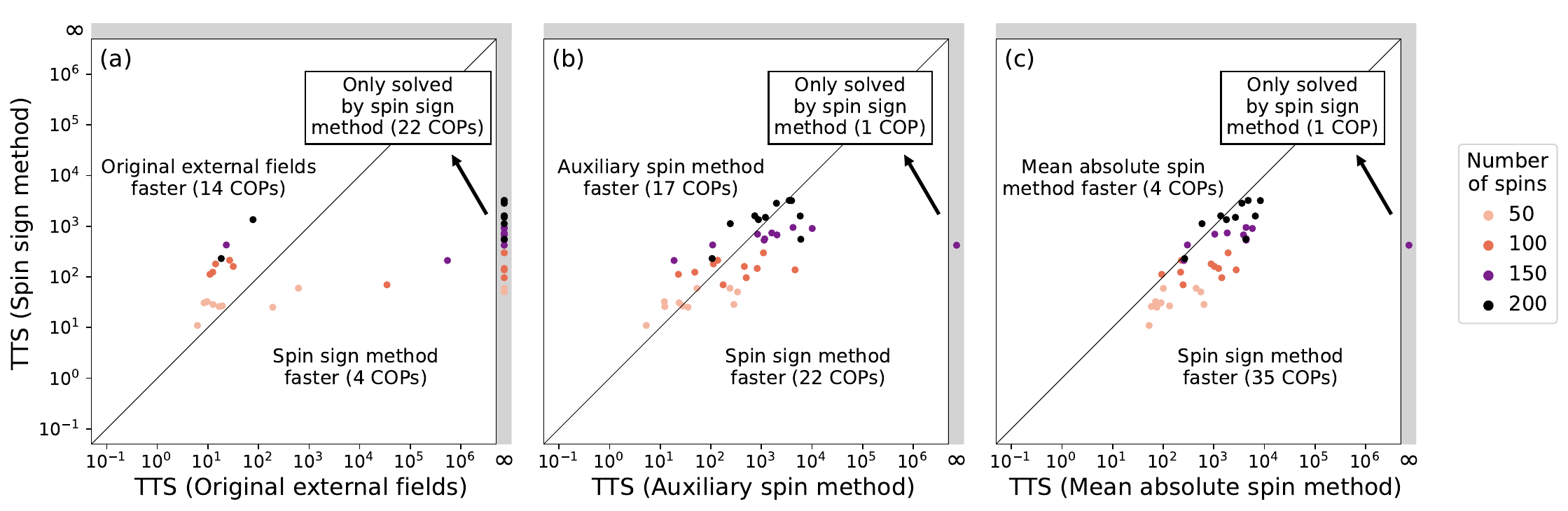}
    \caption{Comparison of time-to-solution between different methods to implement external fields (\cref{eq:original external fields,equation: meanAbs trick,equation: aux trick,eq:spin sign method}) for SK Hamiltonians with random external fields.
    Dots in the grey area on the right denote COPs that could be solved by the spin sign method within the allocated compute time of $t_\text{max}=10^4$, but not by the method on the x-axis ($\text{TTS}=\infty$, $\text{SR}=0$).
    The spin sign method can solve more problems within the allocated time than the other methods, and it generally requires less time to do so.
    }
    \label{fig: SK results: zeta=1}
\end{figure*}

The first type of problem we consider is a Sherrington-Kirkpatrick (SK) spin glass. The goal of this task is to find the configuration of binary spins $\sigma_i \in \{-1, 1\}$ that minimizes the Hamiltonian in \cref{eq:ising hamiltonian binary}, where the elements of the coupling matrix are drawn from a Gaussian distribution \cite{SK_SpinGlass} (with zero mean and standard deviation 1). We extend the SK spin glass model with external fields that are drawn from the same Gaussian distribution.

We considered problems with 50, 100, 150, and 200 spins, generating 10 instances for each size.
For problems with 50 spins, the ground state was obtained via an exact solver \cite{MaxKcut_solver_Fakhimi}. The larger problems could not be solved using this exact solver with our computational resources. Therefore, we used the state-of-the-art approximate solver PySA, which is based on simulated annealing with parallel tempering \cite{PySA}. The parameters used can be found in Supplementary Note 5. The energies obtained by this solver are equal to the lowest energies found by the Ising machine for problems with $100$ and $150$ spins. However, for the problems with $200$ spins, the Ising machine sometimes finds lower energies than PySA. Therefore, we define the lowest energy found by the Ising machine (across all methods of \cref{sec: definition tricks})  as the target energy. Since our primary focus is on the relative comparison of different methods for implementing external fields, this choice has no impact on the validity of our conclusions.

\cref{fig: SK results: zeta=1} compares the various methods of \cref{sec: definition tricks} in terms of TTS when solving SK problems. Each subfigure compares two methods. Each dot on the plot represents a problem. The $x$ ($y$)-coordinate of a dot is the TTS when the problem is solved using the method on the horizontal (vertical) axis. Hence, the diagonal line represents the situation where both methods have the same TTS. 
A dot in the upper left (lower right) of figure denotes a problem that is solved faster by the
method on the horizontal (vertical) axis.

\cref{fig: SK results: zeta=1}(a) compares the original external field method of \cref{eq:original external fields} with the spin sign method of \cref{equation: aux trick}. 22 of the 40 COPs are positioned in the grey region on the right, indicating $\text{TTS}=\infty$ when using the original external fields. This means that for each of these 22 problems and across all tested hyperparameter values, the original external fields failed to reach the target energy in all 100 runs within the maximum time of $t_\text{max}=10^4$ (cf.~\cref{eq:TTS}). In contrast, the spin sign method successfully solved these instances, yielding a finite TTS. The remaining 18 COPs could be solved by either of the two methods. 14 of them were solved faster by the original external fields, while 4 were solved faster using the spin sign method.

In both \cref{fig: SK results: zeta=1}(b) and (c), one of the 40 problems appears in the grey region on the right. Although not visible in the plots, this data point corresponds to the same COP, which could only be solved by the spin sign method, while the other three methods failed.

\cref{fig: SK results: zeta=1}(b) and (c) appear similar, but Fig.~(c) has more dots below the diagonal, indicating that the auxiliary spin method is generally faster than the mean absolute spin method.

In Supplementary Note 10, we provide the values of the success rate and the average runtime $T$ of successful runs corresponding to the TTS values of \cref{fig: SK results: zeta=1}.

Overall, the spin sign method performs best for the SK problems. While it solves all considered COPs, the original external fields fail on more than half. Both the auxiliary spin method and the mean absolute spin method solve all but one COP; however, the latter is generally slower, while the former is only slightly slower than the spin sign method on average, making it the second-best-performing for these problems.

As discussed in \cref{sec:background:zeta-scaling}, previous works proposed to scale external fields by a constant factor $\zeta \in \mathbb{R}$. 
Ref.~\cite{Mean-field_CIM_with_artificial_ZeemanTerms} indicates that the IM's performance, averaged over SK problems with at most 16 spins, is highly sensitive to the choice of $\zeta$. 
However, in the Supplementary Note 6, we show that this sensitivity is due to the small problem size. Our results indicate that applying a scaling factor different from 1 is generally not beneficial here, reinforcing that $\zeta=1$ is the best choice for solving SK problems.
This is explained by the fact that setting $\zeta \neq 1$ removes the guarantee that the ground state spin corresponds to the problem’s optimal solution.

\subsection{Quadratic unconstrained binary optimization problems}
\begin{figure*}[htb]
    \centering
    \includegraphics[width=1.0\linewidth]{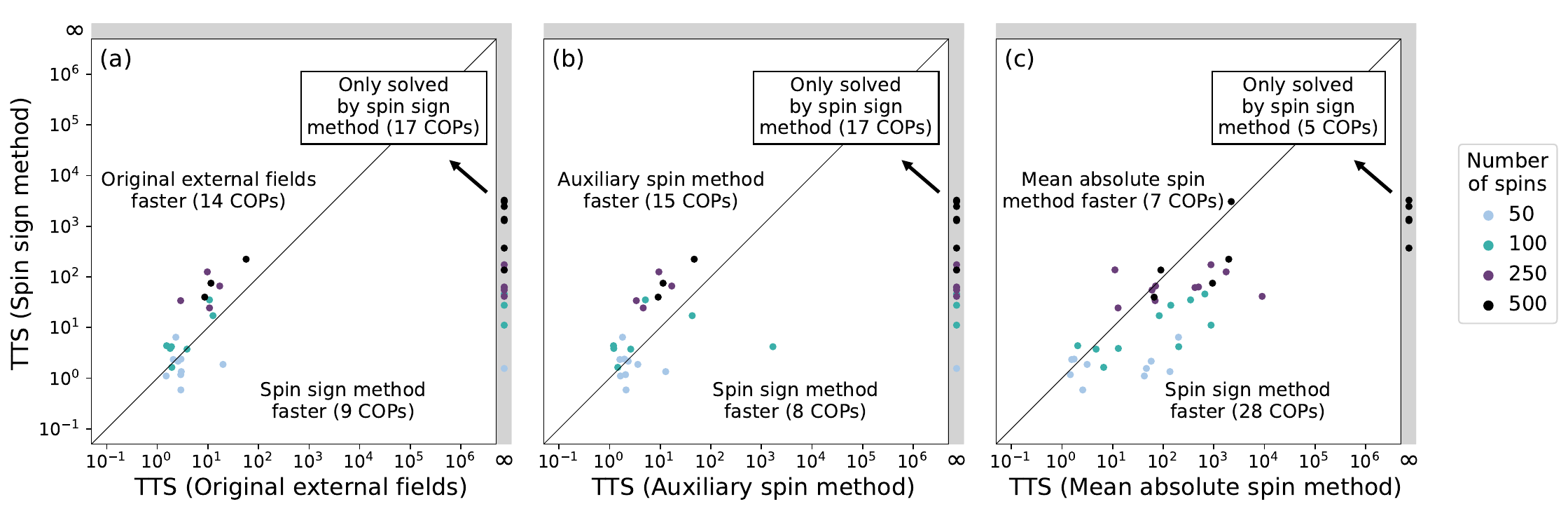}
    \caption{Comparison of time-to-solution between different methods to implement external fields for Beasley problems. 
    Dots in the grey area on the right denote COPs that could be solved by the spin sign method within the allocated compute time of $t_{\text{max}}=10^4$, but not by the method on the x-axis ($\text{TTS}=\infty$, $\text{SR}=0$).}
    \label{fig: beasley: zeta=1}
\end{figure*}
 
The problems analyzed in this section are the Beasley instances from the BiqMac library \cite{BiqMac}.
These are examples of quadratic unconstrained binary optimization (QUBO) problems, meaning that they are defined using binary variables $x_i \in \{0,1\}$. Given a symmetric matrix $Q$, the goal is to find the configuration $\mathbf{x}$ that minimizes the objective function $\text{{\bf x}}^TQ\text{{\bf x}}$.

These problems can be mapped to the Ising model using the following transformation:
\begin{equation}
    x_i = \frac{\sigma_i+1}{2}.
    \label{eq: transfo bits to spin}
\end{equation}

While the original QUBO formulations of these problems include only quadratic terms ($\propto x_i x_j$), Supplementary Note 7 shows that applying \cref{eq: transfo bits to spin} results in both spin couplings ($\propto \sigma_i \sigma_j$) and external fields ($\propto \sigma_i$).
Many COPs of interest are naturally defined as QUBO problems, making the transformation of \cref{eq: transfo bits to spin} a common mechanism that gives rise to external fields in IMs. 

The ground state energies for Beasley problems with up to 250 spins are provided in the BiqMac library \cite{BiqMac}. For problems with 500 spins, the exact ground state is not known. For these problems, we use the upper bounds provided in the BiqMac library \cite{BiqMac} as target values.

\cref{fig: beasley: zeta=1} compares the various methods to incorporate external fields, as defined in \cref{sec: definition tricks}, in terms of TTS, when applied to the Beasley problems. 
\cref{fig: beasley: zeta=1}(a) compares the original external fields of \cref{eq:original external fields} to the spin sign method of \cref{eq:spin sign method}. 17 of the 40 COPs appear at the rightmost edge of the plot, indicating that they could not be solved using the original external fields, while they could be solved using the spin sign method.

\cref{fig: beasley: zeta=1}(b) compares the auxiliary spin method of \cref{equation: aux trick} with the spin sign method of \cref{eq:spin sign method}. The plot looks similar 
to \cref{fig: beasley: zeta=1}(a), and our results indicate that the same 17 COPs at the right edge of \cref{fig: beasley: zeta=1}(a) could also not be solved using the auxiliary spin method.

\cref{fig: beasley: zeta=1}(c) compares the mean absolute spin method of \cref{equation: meanAbs trick} with the spin sign method of \cref{eq:spin sign method}. This time, only 3 COPs of the 40 could be solved exclusively by the spin sign method. Looking at the COPs that could be solved using either of the two methods, which are displayed in the white region of the plot, we see that the spin sign method generally reaches a solution faster.

Overall, we conclude from \cref{fig: beasley: zeta=1} that the spin sign method performs best as it can solve more Beasley problems than the other methods. 
The mean absolute spin method is the second-best-performing method as it solves more COPs than the original external fields and the auxiliary spin method. The latter two methods yield comparable performance.

As explained in \cref{sec:background:zeta-scaling}, it has been proposed in the past to multiply the external fields by a constant factor, but this removes the guarantee that the ground state spin configuration solves these problems. 
In Supplementary Note 6, we demonstrate that applying a factor different from 1 is generally not beneficial for Beasley problems, consistent with our findings for SK problems.

\subsection{Max-3-Cut problems}
\label{sec:results:max3cut}
\begin{figure*}[htb]
    \centering
    \includegraphics[width=1.0\linewidth]{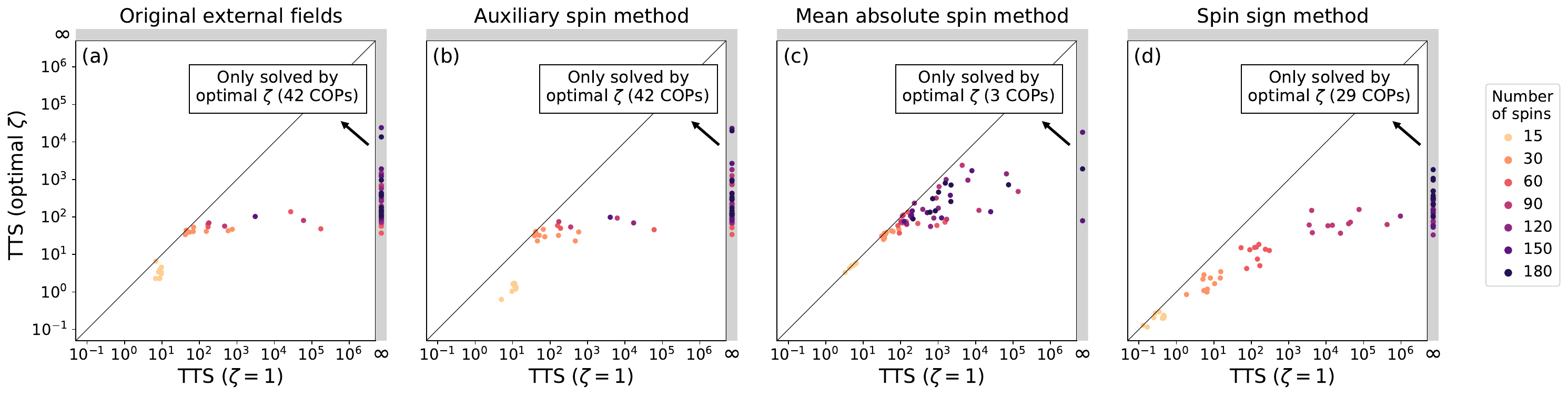}
    \caption{Comparison of time-to-solution between $\zeta=1$ and the optimal $\zeta$ that minimizes the TTS, for each of the methods, for Max-3-Cut problems.
    Data points in the grey area on the right denote COPs that could not be solved using $\zeta=1$ ($TTS=\infty$, $SR=0$), given the allocated compute time of $t_\text{max}=10^4$.}
    \label{fig: max3cut: zeta vs zeta=1}
\end{figure*}

The goal of the Max-3-Cut problem is to partition the vertices of an undirected graph into three sets, maximizing the number of edges that connect different sets, i.e. maximizing the so-called cut-value. In other words, we seek the best possible vertex coloring using 3 colors.
In this section, we consider graphs generated using the rudy generator \cite{rudy_graph_generator}. To benchmark the IM's performance, the problems are solved to optimality using the publicly available \texttt{max\_k\_cut} package \cite{MaxKcut_solver_Fakhimi}.

Max-3-Cut serves as a fundamental example of a COP requiring multivariate integer variables. Such problems are native to the Potts model, an extension of the Ising model, and are typically embedded in an Ising machine using one-hot encoding. This encoding introduces external fields, as will be explained in the next section.

\subsubsection{Ising formulation}
\label{sec:max3cut:mapping}
We utilize the Max-3-Cut Ising formulation as described in Ref. \cite{Ising_formulations_of_many_NP_problems}. 
We denote the sets of vertices and edges in the graph as $V$ and $E$ respectively.
For every vertex $v \in V$, a triplet of binary variables is introduced: $x_{v,i}$ where $i \in \{1,2,3\}$. $x_{v,i}$ equals 1 if vertex $v$ has color $i$, and 0 otherwise. 
The energy is defined as follows:
\begin{equation}
\mathcal{H} = A \sum_{v \in V}\left(1-\sum_{i=1}^3 x_{v, i}\right)^2
+ B \sum_{(uv) \in E} \sum_{i=1}^3 x_{u, i} x_{v, i},
\label{eq:mapping lucas bits}
\end{equation}
where A and B are positive scalars. The first term enforces that all variable triplets $\{x_{v,1}, x_{v,2}, x_{v,3}\}$ are one-hot encoded since this positive term only vanishes if every triplet contains a single 1 and two 0's, ensuring the vertex colors are well-defined. The second term enforces the maximization of the cut value since an energy penalty is added for every edge that connects vertices with the same color. The ratio $B/A$ denotes the relative importance of the two constraints.

The binary variables in \cref{eq:mapping lucas bits} can be transformed to Ising spins using \cref{eq: transfo bits to spin}. As detailed in Supplementary Note 7, this allows us to rewrite the energy as follows:
\begin{equation}
\begin{split}
\mathcal{H} 
=& \frac{A}{4} \sum_v \sum_{i \neq j} \sigma_{v, i} \sigma_{v, j} 
+ \frac{B}{4} \sum_{(uv)\in E} \sum_{i=1}^{3} \sigma_{u, i} \sigma_{v,i} \\
&+ \zeta \sum_v \sum_{i=1}^{3} \left( \frac{A}{2} + \frac{B}{4} \text{deg}(v)\right) \sigma_{v, i},
\end{split}
\label{eq:mapping lucas spins}
\end{equation}
where $\text{deg}(v)$ denotes the degree of vertex $v$. Note that we multiplied the external fields ($\propto\sigma_{v,i}$) by $\zeta \in \mathbb{R}$. Whereas \cref{eq:mapping lucas bits,eq:mapping lucas spins} are only strictly equivalent when $\zeta=1$, we will show in the next section that setting $\zeta \neq 1$ is a necessary modification for the IM to achieve good performance.

\subsubsection{Scaling external fields by a constant counters mapping errors}
\label{sec:max3cut:zeta0.6}

As explained in \cref{sec:background:zeta-scaling}, it is generally not expected that rescaling the external fields with $\zeta \neq 1$ will improve the performance of an IM. Such a rescaling can eliminate the guarantee that the ground-state configuration solves the COP in question. In the previous sections, we confirmed that $\zeta$ is generally best put to 1 when solving an SK problem or Beasley problem.

However, we will show now that this is not the case for the  Max-3-Cut mapping of \cref{eq:mapping lucas spins}. The inspiration for this approach comes from Ref.~\cite{original_meanAbsTrick_paper}, which found that setting $\zeta\approx0.6$ is optimal for a structure-based drug design problem of which the mapping also contains one-hot encoding constraints and edge constraints, similar to \cref{eq:mapping lucas spins}, along with additional external fields.
While the usefulness of this rescaling was observed, its underlying reasons were not explained. 
In this section, we will clarify why this rescaling works for Max-3-Cut.

First, we show that all implementation methods for external fields from \cref{sec: definition tricks} fail when $\zeta=1$, but not when $\zeta\approx0.6$.
For each of these methods, \cref{fig: max3cut: zeta vs zeta=1} compares the case where $\zeta$ is fixed at 1 to the case where $\zeta$ is optimized within $[0,1.2]$. In other words, the x-axis represents $\min_{\alpha,v_\beta,\frac{B}{A}}\text{TTS}$ for $\zeta=1$, while the y-axis represents $\min_{\alpha,v_\beta,\frac{B}{A},\zeta}\text{TTS}$.
As expected, all points lie on or below the diagonal, since tuning $\zeta$ can only maintain or improve performance relative to fixing $\zeta = 1$. Across all four methods, several problems fall into the gray region on the right side of the figure. These instances cannot be solved with $\zeta = 1$ but become solvable when $\zeta$ is optimized within $[0, 1.2]$. Many of the remaining points show that optimizing $\zeta$ can yield improvements in TTS by several orders of magnitude.

\begin{figure}[htb]
    \centering
    \includegraphics[width=1.0\linewidth]{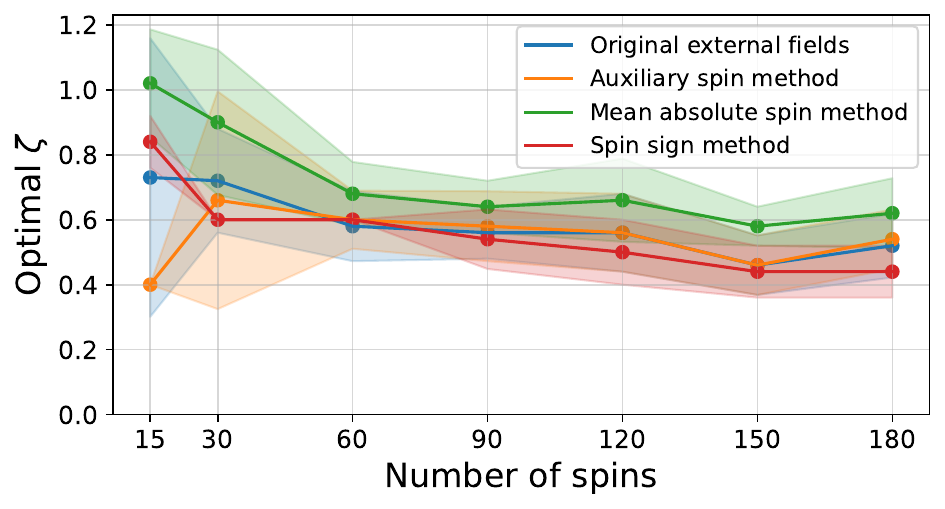}
    \caption{Optimal $\zeta$ values (which minimize TTS) as a function of the number of spins. Each dot represents the mean optimal value across 10 Max-3-Cut problems. Shaded areas represent the standard deviation.
    }
    \label{fig: max3cut: optimal zeta wrt N}
\end{figure}

\cref{fig: max3cut: optimal zeta wrt N} shows the optimal value of $\zeta$ as a function of the graph size. Each data point denotes the average optimal value across 10 problems of the same size. The standard deviation is represented by a shaded area. For all four rescaling methods, the optimal value of $\zeta$ converges to approximately 0.6. Although the optimal $\zeta$ may deviate from 0.6 in some cases, fixing $\zeta=0.6$ generally yields comparable performance to using the instance-specific optimal value, as shown in Supplementary Note 8. This is not the case for other fixed values such as $\zeta=0.4$ or $\zeta=0.8$, which tend to result in worse performance.

This result is consistent with a prior study that identified $\zeta = 0.6$ as the optimal value for a similar problem that also uses one-hot encoding, but with $k>3$ \cite{original_meanAbsTrick_paper} (recall that $k=3$ here). This suggests that the value $\zeta = 0.6$ holds for a broader class of problems using one-hot encoding. 

We now show that the reason why this rescaling is necessary stems from the structure of \cref{eq:mapping lucas spins}. As explained in \cref{sec:background:zeta-scaling}, the ground state configuration of a mapping that contains soft constraints will only solve the COP under consideration when the prefactors of those constraints are set to adequate values. It is often guaranteed that good values for these prefactors exist, but they are not necessarily a priori known \cite{Ising_formulations_of_many_NP_problems}. Consequently, given a value of $B/A$, we can determine the error of the mapping in \cref{eq:mapping lucas spins} as follows. On the one hand, we obtain the \textit{highest achievable cut-value} (the number of edges connecting different sets in the partition) via an exact Max-k-Cut solver \cite{MaxKcut_solver_Fakhimi}. On the other hand, we determine the \textit{cut-value of the ground state} of \cref{eq:mapping lucas spins} via an exhaustive search over the spin configurations. The difference between these respective cut values is defined as the error.

\begin{figure}[htb]
    \centering
    \includegraphics[width=1.0\linewidth]{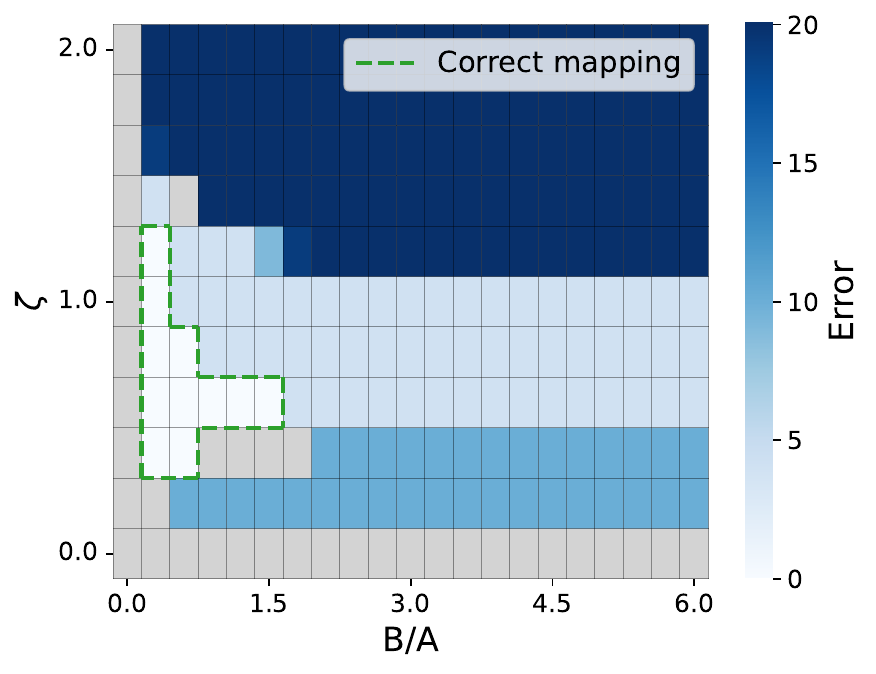}
    \caption{Error of the Max-3-Cut mapping (\cref{eq:mapping lucas spins}) as a function of $\frac{B}{A}$ and $\zeta$ for the g05\_10.0 graph. For every set of values $(\frac{B}{A},\zeta)$, the ground state configuration of \cref{eq:mapping lucas spins} is determined via exhaustive search. The error is defined as the difference between the cut value of this configuration and the optimal cut value (obtained via an exact solver \cite{MaxKcut_solver_Fakhimi}). The region where the ground-state configuration solves the Max-3-Cut problem is surrounded by a green dashed line. Grey cells indicate that the ground state is degenerate, including at least one configuration that does not solve the Max-3-Cut problem.}
    \label{fig: max3cut: correct mapping g05u10d0}
\end{figure}

\cref{fig: max3cut: correct mapping g05u10d0} shows the error of \cref{eq:mapping lucas spins} as a function of $B/A$ and $\zeta$ for an arbitrary Max-3-Cut problem of 30 spins. 
In the white region, outlined by a green dashed line, the ground-state configuration of \cref{eq:mapping lucas spins} correctly solves the COP.
For the blue cells, we verified numerically that the one-hot encoding constraints (the terms of \cref{eq:mapping lucas spins} with prefactor $A$) are violated, leading to ill-defined vertex colors. 
Consequently, the ground-state configuration is not a valid solution in these cells, such that the error is nonzero.

It is important to note that in \cref{fig: max3cut: correct mapping g05u10d0}, the range of valid $B/A$ values is broader for $\zeta=0.6$ compared to $\zeta=1$.
In general, as the graph size increases for $\zeta=1$, it turns out that the mapping remains correct only for progressively smaller values of $B/A$, as larger values violate the one-hot encoding constraints. Interestingly, this issue does not arise for $\zeta=0.6$.
As detailed in Supplementary Note 9, we confirmed these findings for small graphs by evaluating the error of \cref{eq:mapping lucas spins} (requiring an exhaustive search as also used in \cref{fig: max3cut: correct mapping g05u10d0}) and extended them to larger graphs based on the strong correlation between the correctness of the mapping and the success rate.

Although decreasing $B/A$ could theoretically eliminate mapping errors at $\zeta=1$, this would require an unrealistic resolution of the interaction parameters, making the approach impractical. In contrast, choosing $\zeta=0.6$ yields good performance across a much broader range of $B/A$ values, significantly reducing sensitivity to parameter choices. Moreover, as shown in Supplementary Note 8, using $\zeta=0.6$ generally results in similar time-to-solution (TTS) as the optimal $\zeta \in [0,1.2]$. Hence, we conclude that setting $\zeta=0.6$ effectively mitigates mapping errors caused by violations of the one-hot encoding constraint, without requiring impractically small values of $B/A$.

\subsubsection{Comparison of external field implementations}
\label{sec:max3cut:compareTricks}
\begin{figure*}[htb]
    \centering
    \includegraphics[width=1.0\linewidth]{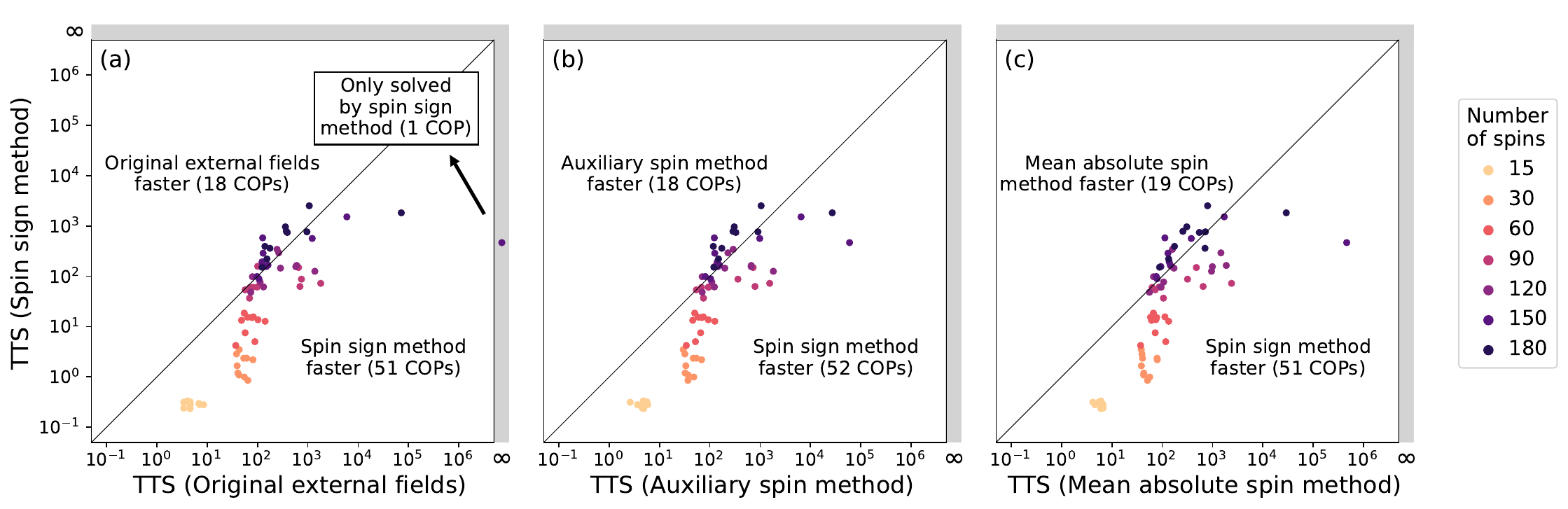}
    \caption{Comparison of time-to-solution between different methods to implement external fields for Max-3-Cut problems for $\zeta=0.6$. The spin sign method generally solves the COPs faster than the other methods. 1 COP could not be solved using the original external fields.}
    \label{fig: max3cut: optimal zeta}
\end{figure*}

\cref{fig: max3cut: optimal zeta} compares the TTS of the methods of \cref{sec: definition tricks} for the Max-3-Cut problems with $\zeta=0.6$. Most data points lie in the lower right of the plots, indicating that the spin sign method solves these COPs faster.
While the advantage over other methods reduces for larger problems, it generally remains the best choice. 
We conclude that the spin sign method also generally performs best for these Max-3-Cut problems, but that the key prerequisite for good performance is setting $\zeta \approx 0.6$.

\section{Hardware considerations}
In the previous section, we have shown that the spin sign method is the most effective approach to incorporate external fields in analog IMs. Additionally, it can easily be implemented in hardware using, for example, a comparator in electronic systems. Moreover, analog spin amplitudes are often measured before computing the local fields $I_i$ \cite{2000nodes,100000SpinsCIM,Poor_mans_CIM,Ising_machine_based_on_networks_of_subharmonic_electrical_resonators,jiang2023efficient,PolaritonCondensates,PolaritonCondensates_2}. In such setups, a one-bit resolution measurement device can enforce the spin sign method, provided this only affects the local field computation and preserves the analog nature of the spins.

Whereas the spin sign method offers the best performance and can be implemented in hardware relatively easily, the mean absolute spin method performs worse and further complicates the calculation of $I_i$. Indeed, in addition to the standard matrix-vector multiplication, it requires computing the mean absolute spin value, which will require extending the hardware.

The auxiliary spin method, though less effective than the spin sign method, may still be useful when directly implementing external fields in hardware is challenging. However, since the auxiliary spin must connect to all spins affected by an external field, it can pose difficulties in hardware that does not allow for all-to-all coupling. Moreover, while this method is often cited as a justification for excluding external fields from hardware implementations, we emphasize that this approach is not ideal, as it leads to suboptimal performance.

Hence, the spin sign method not only provides the best performance but is also relatively easy to implement compared to the auxiliary spin and mean absolute spin methods.

\section{Conclusion}
The initial implementation of external fields in IMs with analog spins, \cref{eq:original external fields}, is prone to imbalances in magnitude between the external fields and the spin couplings. To address this, the mean absolute spin method of \cref{equation: meanAbs trick} and the auxiliary spin method of \cref{equation: aux trick} were proposed in the past. 

In this work, we demonstrate that the spin sign method of \cref{eq:spin sign method} consistently outperforms the earlier approaches across three distinct problem classes. For SK and Beasley instances, it enables solving more COPs within the allotted time, while for Max-3-Cut, it generally achieves faster solutions than any of the other methods.

Although the spin sign method was previously shown to be effective for ballistic simulated bifurcation \cite{inspiration_idea_thomas,higher_order_dSB}, its use has remained largely confined to that setting, where $\mathcal{F}$ in \cref{eq:generic_update_eq} is linear and spin dynamics involve momentum. It has not been widely adopted in other analog IMs \cite{OverviewMcMahon}, nor included in prior studies on balancing spin couplings and external fields \cite{Mean-field_CIM_with_artificial_ZeemanTerms}.
In this work, we extend its application and show that it outperforms standard approaches for handling magnitude imbalances between spin couplings and external fields.

Beyond refining IM dynamics, we shed light on the embedding process of certain COPs. Contrary to some prior works, we showed that the external fields should not be rescaled with a constant $\zeta \neq 1$ for problems without soft constraints, such as SK and Beasley problems. However, we confirmed that such a rescaling can be necessary when soft constraints are used, as in Max-3-Cut. While this necessity had been observed before, we identified the underlying reason: with $\zeta = 1$, one-hot encoding constraints are often violated due to the finite resolution of interaction parameters. Choosing $\zeta \approx 0.6$ restores the embedding's correctness for the problems considered here. Notably, prior work also found this value to be optimal for one-hot encoding with more than three states \cite{original_meanAbsTrick_paper}, suggesting its robustness. Since many COPs represent multivariate variables using one-hot encoding, this insight is valuable for the Ising machine community. Moreover, since real-world COPs often involve numerous soft constraints—particularly when using IMs—it may be worthwhile to explore whether similar rescalings could benefit other types of soft constraints.

\section{Data availability}
The authors declare that all relevant data are included in the manuscript. Additional data are available from the corresponding author upon reasonable request.

\section{Author contributions}
R.D.P. and J.L. performed the simulations and wrote the manuscript. P.B., G.V.d.S., G.V. and T.V.V. supervised the project. All authors discussed the results and reviewed the manuscript.

\section{Additional information}
{\bf Competing interests:} Guy Van der Sande is an Editorial Board Member for Communications Physics, but was not involved in the editorial review of, or the decision to publish this article. All the other authors declare no competing interests.\\
{\bf Acknowledgements:} 
The authors would like to thank Ruqi Shi for generating graphs that were used to study Max-3-Cut, and Toon Sevenants for the many interesting discussions.

This research is funded by the Prometheus Horizon Europe project 101070195.
It was also funded by the Research Foundation Flanders (FWO) under grants G028618N, G029519N, G0A6L25N and G006020N. Additional funding was provided by the EOS project `Photonic Ising Machines'. This project (EOS number 40007536) has received funding from the FWO and F.R.S.-FNRS under the Excellence of Science (EOS) programme.

\bibliographystyle{unsrtnat}
\bibliography{references}
\clearpage



\renewcommand{\theequation}{S.\arabic{equation}}
\renewcommand{\thetable}{S\arabic{table}}
\renewcommand{\thefigure}{S\arabic{figure}}
\setcounter{equation}{0}
\setcounter{figure}{0}

\twocolumn[
\vspace*{1cm}
\begin{center}
  {\LARGE \bfseries Supplementary Material}
\end{center}
\vspace{0.5cm}
]

\section*{Supplementary Note 1: Simulation of the Ising machine}
Simulating a run of the Ising machine begins by drawing the initial spin amplitudes from the same Gaussian distribution as the noise (see further). Their amplitudes are updated each time step according to the Euler scheme:
\begin{equation}
    \label{eq:EulerScheme}
    \text{{\bf s}}_{t+1} = \text{{\bf s}}_t + \Delta t \left(-\text{{\bf s}}_t + \tanh\big(\alpha \text{{\bf s}}_t + \beta_t \text{{\bf I}}_t\big)\right) + \gamma \boldsymbol{\xi}_t,
\end{equation}
where $\text{{\bf s}}_t$ is the vector of all spin amplitudes at time $t$, $\Delta t = 0.01$ is the Euler time step, $\alpha$ is the linear gain, $\text{{\bf I}}_t$ is the vector of the local fields of all spin amplitudes at time $t$, $\gamma=0.001$ is the noise strength and $\boldsymbol{\xi}_t$ is the vector of the noise terms of all spin amplitudes. Each noise term is drawn from a Gaussian distribution with zero mean and a standard deviation of $\sqrt{\Delta t}$. $\beta_t$ is the interaction strength at time $t$ and is given by
\begin{equation}
    \label{eq:LinearAnnealing_suppl}
    \beta_{t+1} = \beta_t + v_\beta\Delta t,
\end{equation}
where $v_\beta$ is the annealing speed. The initial $\beta$ value is $\beta_0$ and depends on the method used to implement the external fields. For the original external fields and the spin sign method, $\beta_0=0$ is used as the origin, i.e. all spin amplitudes equal to zero, is a fixed point only at $\beta=0$. For the auxiliary spin method, $\beta_0$ is set to the value at which the origin undergoes its first pitchfork bifurcation, as described in Supplementary Note 4. Because computing this bifurcation point when using the mean absolute spin method is computationally expensive, we also set $\beta_0=0$ in that case. More details can be found in Supplementary Note 3.

The updates in \cref{eq:EulerScheme} and \cref{eq:LinearAnnealing_suppl} are repeated until either the IM obtains the target energy or $10^4/\Delta t$ updates are done.

For each problem class, the TTS was calculated for all possible combinations of parameter values within the ranges specified in \cref{tab:OptimalParameters} via a grid search. The number of points per parameter in the last column determines the resolution of the sweep. The TTS mentioned in the main text is the lowest TTS obtained in this way. 

As discussed in the main text, Max-3-Cut problems involve an additional parameter, $B/A$, which controls the relative importance of the edge constraints (weighted by $B$) compared to the one-hot encoding constraints (weighted by $A$). The term proportional to $A$ is a sum over all vertices and therefore scales with the number of vertices $N_\text{vert}$, while the term proportional to $B$ scales with the number of edges, i.e., $\propto N_\text{vert}^2$. Consequently, the same relative importance between constraints is achieved at lower values of $B/A$ as $N_\text{vert}$ increases. Since we use a fixed number of sampling points ($7$) in our parameter scan, we rescale the range of $B/A$ with $1/N_\text{vert}$ to ensure that the scanned values remain relevant and adequately capture the region of interest for each value of $N_\text{vert}$.

\begin{table*}[!ht]
\caption{\textbf{Parameter sweeps.} The ranges of the hyperparameters used to determine their optimal values.}
\centering
\renewcommand{\arraystretch}{1.2}
\begin{tabular}{|p{0.12\linewidth}|p{0.12\linewidth}|p{0.11\linewidth}|p{0.11\linewidth}|p{0.13\linewidth}|p{0.1\linewidth}|}
\hline
\textbf{Problem class} & \textbf{Parameter name} & \textbf{Lowest value} & \textbf{Highest value} & \textbf{Spacing type} & \textbf{Number of points} \\
\hline\hline
\multirow{3}{*}{SK} 
    & $\alpha$   & $-10$       & $1$         & Linear           & 12 \\
    & $v_\beta$  & $10^{-5}$   & $10^{-1}$   & Log (base 10)    & 5 \\
    & $\zeta$    & $0$         & $2$         & Linear           & 11 \\
\hline
\multirow{3}{*}{Beasley} 
    & $\alpha$   & $-10$       & $1$         & Linear           & 12 \\
    & $v_\beta$  & $10^{-5}$   & $10^{-1}$   & Log (base 10)    & 5 \\
    & $\zeta$    & $0$         & $2$         & Linear           & 11 \\
\hline
\multirow{4}{*}{Max-3-Cut} 
    & $\alpha$   & $-10$       & $1$         & Linear           & 5 \\
    & $v_\beta$  & $10^{-5}$   & $10^{-1}$   & Log (base 10)    & 5 \\
    & $\zeta$    & $0$         & $1.2$       & Linear           & 7 \\
    & $B/A$      & $0$         & $60/N_\text{vert}$      & Linear           & 7 \\
\hline
\end{tabular}
\label{tab:OptimalParameters}
\end{table*}

\section*{Supplementary Note 2: Initial search direction without rescaling of the external fields}
When using the original external fields method of Eq.~3 of the main paper, the time evolution of the spin amplitudes when using the tanh nonlinearity is given by 
\begin{equation}
    \label{eq:Supp_tanh}
    \frac{ds_i}{dt} = - s_i + \tanh\left(\alpha s_i +  \beta\Big(\sum_j J_{ij} s_j + h_i\Big)\right).
\end{equation}
When $\beta = 0$, the origin (all spin amplitudes equal to zero) is a fixed point of the system. We are now interested in the state that emerges when $\beta$ is slightly increased. This means that at a fixed point, for all $i$
\begin{equation}
\label{eq:beta}
    \beta = \frac{-\alpha s_i + \tanh^{-1}(s_i)}{\sum_j J_{ij} s_j + h_i}.
\end{equation}
The partial derivative with respect to $s_k$, evaluated at the origin is given by
\begin{equation}
    \label{eq:dbeta_ds}
    \frac{\partial\beta}{\partial s_k}\biggr\rvert_{\text{{\bf s}}=\boldsymbol{0}, \beta=0} = \frac{1-\alpha}{h_k}.
\end{equation}
Since \cref{eq:beta} is invertible for $\alpha < 1$, we can write:
\begin{equation}
    \label{eq:ds_dbeta}
    \frac{\partial s_k}{\partial\beta}\biggr\rvert_{\text{{\bf s}}=\boldsymbol{0}, \beta=0} = \frac{h_k}{1-\alpha}.
\end{equation}
This means that, when starting from the origin and increasing $\beta$, the system will evolve in a direction proportional to the external field $h_i$, independent of the values of the spin couplings $J_{ij}$.

\section*{Supplementary Note 3: Bifurcation of the mean absolute spin method}

In general, the stability of a fixed point is related to the eigenvalues of the Jacobian matrix evaluated at that fixed point. When all eigenvalues are negative, the fixed point is stable. When one eigenvalue becomes positive, the fixed point becomes unstable in the direction of the corresponding eigenvector.

When using the mean absolute spin method in combination with the tanh nonlinearity, the evolution of the spin amplitudes is given by:
\begin{equation}
    \label{eq:SuppMeanAbsSigmoid}
    \frac{ds_i}{dt} = -s_i + \tanh\bigg(\alpha s_i + \beta\Big(\sum_j J_{ij} s_j + h_i\frac{1}{N}\sum_k|s_k|\Big)\bigg).
\end{equation}
The origin bifurcates when its stability changes, i.e. when the largest eigenvalue of the Jacobian matrix goes through zero and becomes positive. The Jacobian matrix evaluated at the origin is given by 
\begin{multline}
    \label{eq:JacobianMeanAbs}
    \mathcal{U}_{il}\biggr\rvert_{\text{{\bf s}}=\boldsymbol{0}} = -\frac{\partial \frac{ds_i}{dt}}{\partial s_l}\biggr\rvert_{\text{{\bf s}}=\boldsymbol{0}} \\= (\alpha - 1)\delta_{il} + \beta J_{il} + \beta \frac{h_l}{N}\lim_{s_l \rightarrow 0}\text{sign}(s_l),
\end{multline}
where $\delta_{il}$ is the Kronecker delta. So, the Jacobian matrix when using the mean absolute spin method is discontinuous at the origin. This means that, in each of the $2^N$ directions around the origin, the Jacobian has different eigenvalues and eigenvectors. Therefore, to locate the bifurcation point of the origin, one must substitute the signs of the spins of all directions in Eq.~\eqref{eq:JacobianMeanAbs} and determine whether the principal eigenvector has the same signs. The origin will bifurcate in the direction of such an eigenvector with the eigenvalue that first becomes positive. Because of this, determining the $\beta$ value of the bifurcation point of the origin when using the mean absolute spin method is computationally more demanding than finding the ground state by a brute force search, which defeats the purpose of using an Ising machine.

As the bifurcation point cannot be easily found, the annealing technique is initialized at $\beta_0 = 0$. At this point, however, the origin is still stable in all directions and the spin amplitudes will just remain at zero. The spin amplitudes will only deviate from zero when $\beta$ has been annealed to a value high enough so that the origin is unstable in at least one direction. This means that the inability to analytically calculate the bifurcation point of the mean absolute spin method leads to wasted simulation time and therefore a slightly inflated TTS.

While it is, in principle, possible to develop a heuristic to estimate an approximate $\beta_0$ for the mean absolute spin method—potentially improving its performance—we do not pursue that direction here. Instead, we consider a complementary scenario: setting $\beta_0 = 0$ for the auxiliary spin method, rather than using the value defined in \cref{eq:beta_0 aux} (the latter equation is used to obtain the results in the main text). This intentionally introduces an amount of wasted simulation time, similar to that of the mean absolute spin method, allowing us to assess the impact of $\beta_0$ on the TTS of the IM.

\begin{figure*}[htb]
    \centering
    \includegraphics[width=1.0\linewidth]{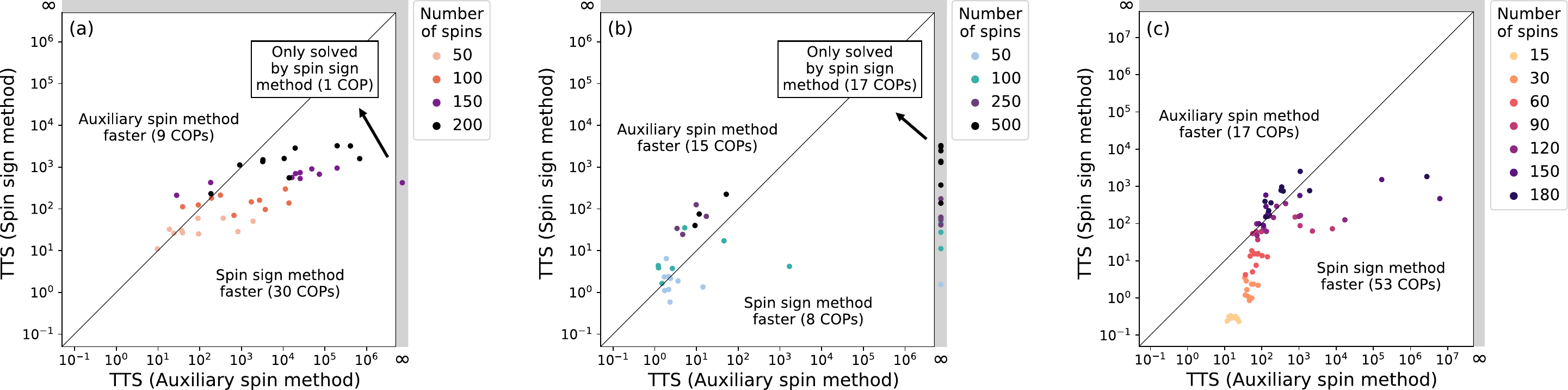}
    \caption{Comparison of time-to-solution between the auxiliary spin method and the spin sign method for (a) SK problems, (b) Beasley problems, and (c) Max-3-Cut problems. These plots replicate Figures 2(b), 3(b), and 7(b) from the main text, but with the auxiliary spin method initialized at $\beta_0 = 0$ instead of the value given in \cref{eq:beta_0 aux}. Comparison with the original figures shows that the choice of $\beta_0$ has a limited effect on the TTS and does not alter the relative performance between methods.
    }
    \label{fig: aux where beta_0=0}
\end{figure*}

\cref{fig: aux where beta_0=0} illustrates this scenario for (a) SK problems, (b) Beasley problems, and (c) Max-3-Cut problems. It mirrors Figures 2(b), 3(b), and 7(b) from the main text, but with the data shifted horizontally to the right to reflect the new initialization at $\beta_0 = 0$ for the auxiliary spin method. Comparing \cref{fig: aux where beta_0=0} with the original figures, we find that this shift modestly affects the SK problem data, while its impact on the Beasley and Max-3-Cut problems is negligible. Importantly, even for the SK problems, the shift is small enough that it does not alter the relative performance between methods shown in Fig.~2 of the main text. From these observations we conclude that setting $\beta_0  = 0$ for the mean absolute spin method is not the main cause of the lower performance of this method.

\section*{Supplementary Note 4: Bifurcation of the auxiliary spin method}
When using the auxiliary spin method in combination with the tanh nonlinearity, the evolution of the spin amplitudes is given by:
\begin{equation}
    \label{eq:SuppAuxSigmoid}
    \frac{ds_i}{dt} = -s_i + \tanh\bigg(\alpha s_i + \beta\sum_j J'_{ij} s_j\bigg),
\end{equation}
where $\mathbf{J'}$ is the extended interaction matrix introduced in section 2.3.3 of the main text. The Jacobian matrix evaluated at the origin is given by 
\begin{equation}
    \label{eq:JacobianAux}
    \mathcal{U}_{il}\biggr\rvert_{\text{{\bf s}}=\boldsymbol{0}} = -\frac{\partial \frac{ds_i}{dt}}{\partial s_l}\biggr\rvert_{\text{{\bf s}}=\boldsymbol{0}} = (\alpha - 1)\delta_{il} + \beta J'_{il}.
\end{equation}
The eigenvalues of this matrix $\lambda_i$ are given by
\begin{equation}
    \label{eq:EigenvaluesAux}
    \lambda_i = \alpha - 1 + \beta \mu_i,
\end{equation}
where $\mu_i$ are the eigenvalues of the extended interaction matrix $\mathbf{J'}$. When all eigenvalues $\lambda_i$ are negative, the origin is stable. When one eigenvalue $\lambda_i$ becomes positive, the origin becomes unstable in the direction of the corresponding eigenvector. This happens when
\begin{gather}
       \label{eq:BifurcationtPointAux}
    \lambda_+ = 0 = \alpha - 1 + \beta \mu_+\\
    \beta = \frac{1-\alpha}{\mu_+}, 
\label{eq:beta_0 aux}
\end{gather}
where $\mu_+$ is the largest eigenvalue of the extended interaction matrix $\mathbf{J'}$.

\section*{Supplementary Note 5: PySA operation}
The PySA algorithm performs simulated annealing with parallel tempering on $n_{\text{replicas}}$ replicas at different temperatures. These replicas are initialized in random states and at a temperature given by
\begin{equation*}
    T_k = \begin{cases}
        T_\text{min} R^{k-1}& k=1, ..., n_{\text{replicas}}-1\\
        0& k=n_{\text{replicas}}
    \end{cases}
\end{equation*}
where $T_k$ is the temperature of replica $k$ and $R$ is given by
\begin{equation*}
    R = \text{exp}\Bigg(\frac{1}{n_{\text{replicas}}-2}\log\left(\frac{T_\text{max}}{T_\text{min}}\right)\Bigg).
\end{equation*}
So, there is one replica at temperature zero and the other replicas have temperatures distributed between $T_\text{min}$ and $T_\text{max}$. 

The replicas are then updated in sweeps. In our case, a sweep consists of attempting a spin flip for each spin in random order. In total $n_{\text{sweeps}}$ are performed for each replica. 

This procedure of initializing and performing sweeps is repeated $n_{\text{reads}}$ times.

In order to search for the ground state energies of the Sherrington-Kirkpatrick (SK) spin glasses, the algorithm was run five times, using different values for the parameters. The values of these parameters can be found in \cref{tab:PySA_parameters}. This resulted in the same energies that were obtained by the Ising machine simulations for problems of $100$ and $150$ spins. However, for the problems with $200$ spins, the Ising machine found equal or lower energies than PySA.

\begin{table}[!ht]
\caption{\textbf{PySA parameters.} Parameter values used to search for the ground state energy of SK problems.}
\centering
\renewcommand{\arraystretch}{1.2}
\begin{tabular}{|c|c|c|c|c|c|}
\hline
\textbf{Run} & $n_{\text{sweeps}}$ & $n_{\text{replicas}}$ & $n_{\text{reads}}$ & $T_{\text{min}}$ & $T_{\text{max}}$ \\
\hline\hline
1 & 100   & 5000 & 1000 & 0.01  & 3.0  \\
2 & 1000  & 500  & 1000 & 0.01  & 3.0  \\
3 & 500   & 500  & 1000 & 0.001 & 6.0  \\
4 & 500   & 1000 & 500  & 0.001 & 10.0 \\
5 & 500   & 1000 & 800  & 0.001 & 15.0 \\
\hline
\end{tabular}
\label{tab:PySA_parameters}
\end{table}

\section*{Supplementary Note 6: Optimal $\boldsymbol{\zeta}$ for SK and Beasley problems}
\begin{figure*}[htb]
    \centering
    \includegraphics[width=1.0\linewidth]{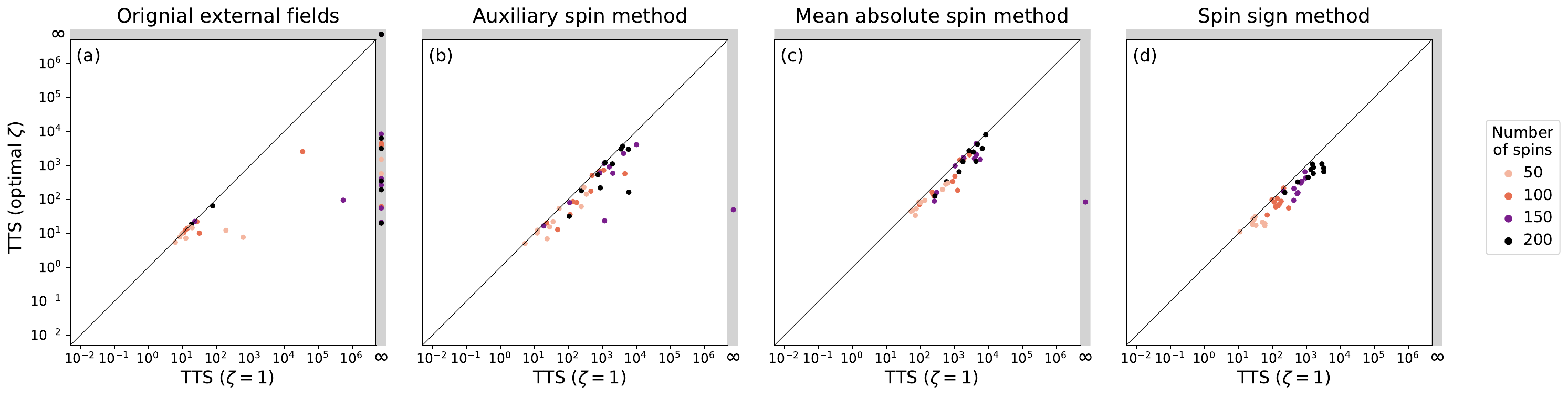}
    \caption{Comparison of time-to-solution between $\zeta=1$ and the optimal $\zeta$ that minimizes the TTS, for each of the methods, for SK problems.
    In Fig.~(a), 6 data points coincide in the grey area in the top right. The corresponding COPs could not be solved with $\zeta=1$, nor with optimal $\zeta$ (i.e. $TTS=\infty$, $SR=0\%$), given the allocated compute time of $t=10^4$. 
    }
    \label{fig: SK: zeta vs zeta=1}
\end{figure*}

\begin{figure*}[htb]
    \centering
    \includegraphics[width=1.0\linewidth]{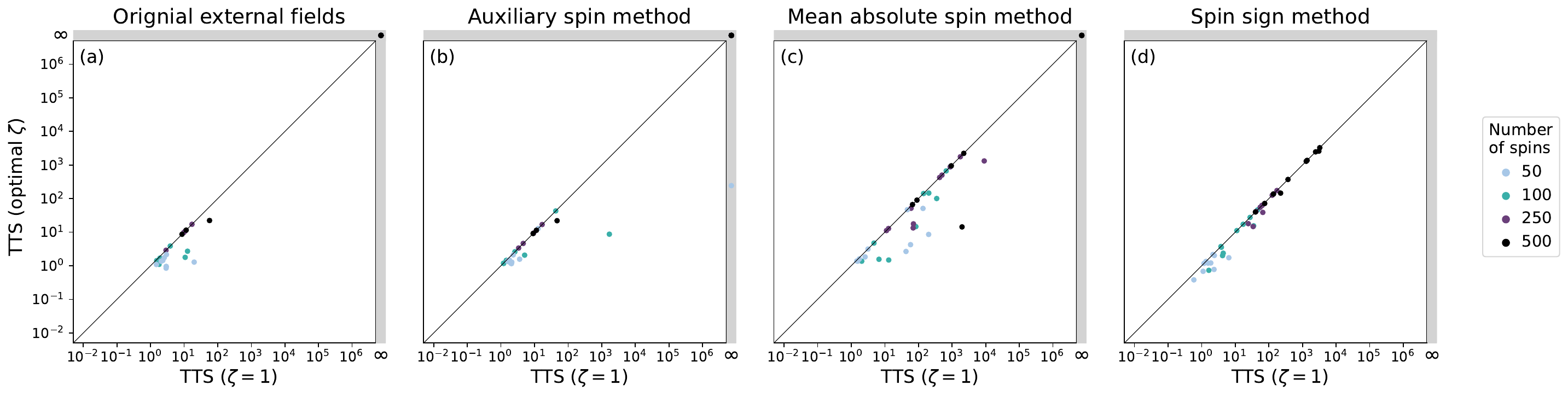}
    \caption{{\bf Comparison of time-to-solution between $\boldsymbol{\zeta=1}$ and the optimal $\boldsymbol{\zeta}$ that minimizes the TTS, for each of the methods, for Beasley problems.}
    If COPs could not be solved with $\zeta=1$, nor with optimal $\zeta$ (i.e. $TTS=\infty$, $SR=0\%$), given the allocated compute time of $t=10^4$, then they appear in the grey area at the top right. 
    These data points coincide. The number of such COPs in Fig.~(a)–(d) is 17, 16, 5, and 0, respectively.}
    \label{fig: beasley: zeta vs zeta=1}
\end{figure*}

As discussed in section 2 of the main text, setting $\zeta=1$ guarantees that the ground state of the binary Hamiltonian solves the original COP. This guarantee is lost for $\zeta\neq1$. However, small deviations from $\zeta=1$ can still be expected to provide valid solutions in many cases. Moreover, adding $\zeta$ as an additional tunable parameter cannot deteriorate the performance. \cref{fig: SK: zeta vs zeta=1,fig: beasley: zeta vs zeta=1} examine whether such improvements are practically meaningful. \cref{fig: SK: zeta vs zeta=1} compares the TTS obtained using an optimized $\zeta$ with the TTS using fixed $\zeta=1$ for SK problems; \cref{fig: beasley: zeta vs zeta=1} shows the same comparison for Beasley instances. In both cases, when methods other than the original external fields are used, most data points lie close to the diagonal, indicating that optimizing $\zeta$ leads to minimal improvement in TTS. This effect is most pronounced for the spin sign method.

\section*{Supplementary Note 7: Transformation from binary spins to bipolar spins}
When solving the Beasley instances of the BiqMac library, the goal is to find the configuration of binary spins $x_i \in \{0,1\}$ that minimizes $\text{{\bf x}}^TQ\text{{\bf x}}$, with $Q$ a symmetric matrix.

This problem can be rewritten in terms of bipolar spins $\sigma_i\in\{-1,1\}$ using the transformation
\begin{equation}
    x_i = \frac{\sigma_i+1}{2}.
    \label{eq: Suppl transfo binary spins to spin}
\end{equation}
The objective function now becomes
\begin{align}
    \sum_{ij}Q_{ij}x_ix_j &= \sum_{ij}Q_{ij}\frac{\sigma_i+1}{2}\frac{\sigma_j+1}{2}\\
    &= \frac{1}{4}\sum_{ij}Q_{ij}\Big(\sigma_i\sigma_j + \sigma_i + \sigma_j + 1\Big).
\end{align}
Using that $Q$ is symmetric
\begin{equation}
    \sum_{ij}Q_{ij}\sigma_i = \sum_{ij}Q_{ji}\sigma_j = \sum_{ij}Q_{ij}\sigma_j,
\end{equation}
We can rewrite the objective function with the following definitions: $J_{ij}:=-\frac{1}{2}\tilde{Q}_{ij}$ and $h_i:=-\frac{1}{2}\sum_j Q_{ij}$, where $\mathbf{\tilde{Q}}$ is the same as $\mathbf{Q}$, but with the diagonal elements set to zero.
\begin{equation}
    -\frac{1}{2}\sum_{ij}J_{ij}\sigma_i \sigma_j - \sum_i h_i\sigma_i.
\end{equation}
Here, we neglected the constant term and the diagonal elements of the interaction matrix, as they are of no importance when minimizing the objective function.

Note that the problem in terms of binary spins does not have any external fields, but the same problem in terms of bipolar spins does. 

When tackling Max-3-Cut problems, a similar procedure can be applied. The Hamiltonian of such problems is given by 
\begin{equation}
\mathcal{H} = A \sum_{v \in V}\left(1-\sum_{i=1}^3 x_{v, i}\right)^2
+ B \sum_{(uv) \in E} \sum_{i=1}^3 x_{u, i} x_{v, i},
\label{eq:Max3Cut_Lucas}
\end{equation}
This can again be rewritten in terms of bipolar spins by using the transformation of \cref{eq: Suppl transfo binary spins to spin}.
\begin{multline*}
    \mathcal{H} = A \sum_{v \in V}\left(1-\sum_{i=1}^3 \frac{\sigma_{v, i}+1}{2}\right)^2\\ 
+ B \sum_{(uv) \in E} \sum_{i=1}^3 \frac{\sigma_{u, i}+1}{2} \frac{\sigma_{v, i}+1}{2}.
\end{multline*}
We can rewrite the first term as follows:
\begin{multline*}
    A \sum_{v \in V}\left(1-\sum_{i=1}^3 \frac{\sigma_{v, i}+1}{2}\right)^2 \\
    = A \sum_{v \in V}\Bigg(1+\bigg(\sum_{i=1}^3 \frac{\sigma_{v, i}+1}{2}\bigg)^2 - 2 \sum_{i=1}^3 \frac{\sigma_{v, i}+1}{2}\Bigg) \\
    = A \sum_{v \in V}\Bigg(\sum_{i,j=1}^3 \frac{\sigma_{v, i}\sigma_{v, j}}{4} + \sum_{i=1}^3 \frac{\sigma_{v, i}}{2} + \frac{1}{4}\Bigg).
\end{multline*}
The second term can be rewritten as
\begin{align*}
    &B \sum_{(uv) \in E} \sum_{i=1}^3 \frac{\sigma_{u, i}+1}{2} \frac{\sigma_{v, i}+1}{2}\\
    &= B \sum_{(uv) \in E} \sum_{i=1}^3 \left(\frac{\sigma_{u, i} \sigma_{v, i}}{4} + \frac{1}{4}\right) \\
    &\quad + B \sum_{(uv) \in E} \sum_{i=1}^3 \left( \frac{\sigma_{u, i}}{4} + \frac{\sigma_{v, i}}{4} \right) \\
    &= B \sum_{(uv) \in E} \sum_{i=1}^3 \left( \frac{\sigma_{u, i} \sigma_{v, i}}{4} + \frac{1}{4} \right) \\
    &\quad + B \sum_{v \in V} \sum_{i=1}^3 \frac{\sigma_{v, i}}{4} \deg(v)
\end{align*}
where in the last line, we changed the sum of the second term from going over all edges of the graph, to going over all vertices. This, while once again ignoring all constant terms gives 
\begin{equation}
\begin{split}
\mathcal{H} 
=& \frac{A}{4} \sum_{v \in V} \sum_{i \neq j} \sigma_{v, i} \sigma_{v, j} 
+ \frac{B}{4} \sum_{(uv)\in E} \sum_{i=1}^{3} \sigma_{u, i} \sigma_{v,i} \\
&+ \sum_{v \in V} \sum_{i=1}^{3} \left( \frac{A}{2} + \frac{B}{4} \text{deg}(v)\right) \sigma_{v, i},
\end{split}
\end{equation}

\section*{Supplementary Note 8: Optimal $\boldsymbol{\zeta}$ for Max-3-Cut problems}
\begin{figure*}[htb]
    \centering
    \includegraphics[width=1.0\linewidth]{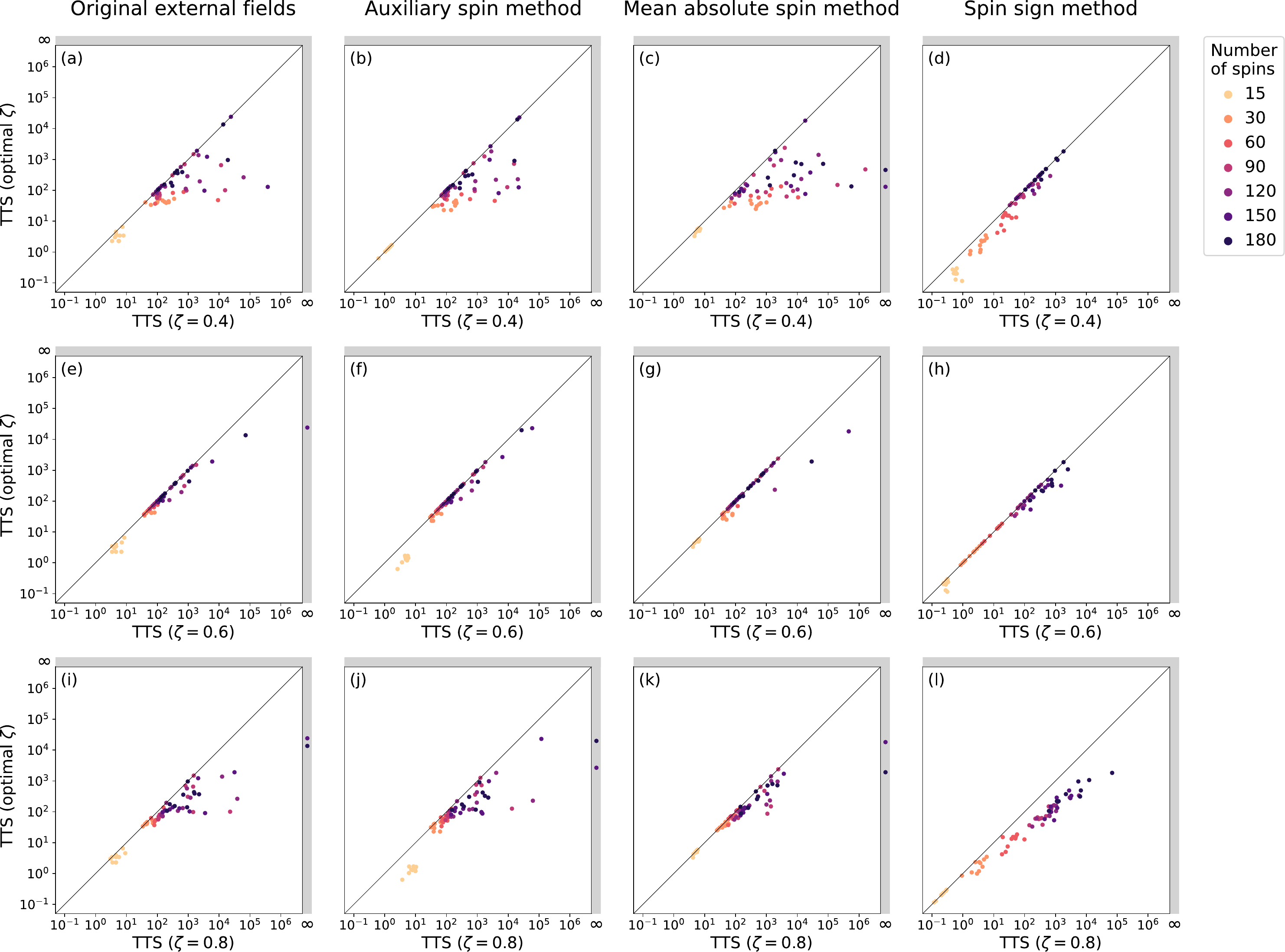}
    \caption{{\bf Comparison of the TTS achieved with optimized $\boldsymbol{\zeta}$ values versus fixed $\boldsymbol{\zeta \in \{0.4,0.6,0.8\}}$ for Max-3-Cut instances.} Each column corresponds to one of the four methods to implement the external fields. The vertical axis always shows the TTS obtained using the optimal $\zeta$ per instance, while the horizontal axis shows the TTS with a fixed $\zeta \in \{0.4,0.6,0.8\}$. Each point represents a single problem instance. Points lying on the diagonal indicate equal performance between fixed and optimized $\zeta$.}
    \label{fig: max3cut: optimal zeta vs constant zeta}
\end{figure*}

As explained in the main text, the optimal $\zeta$ value for Max-3-Cut problems converges to 0.6 as the problem size increases. While there is some variation on this problem-specific optimal $\zeta$ value, using $\zeta=0.6$ yields comparable performance across all problems and methods of implementing the external fields. \cref{fig: max3cut: optimal zeta vs constant zeta} compares the time-to-solution (TTS) achieved using the optimized value of $\zeta$ (vertical axis) versus a fixed value of $\zeta$ (horizontal axis) for all Max-3-Cut problem instances. Each row corresponds to a different fixed value: $\zeta=0.4$, $\zeta=0.6$, and $\zeta=0.8$ and each column represents one of the four methods considered to implement external fields. In these plots, each point represents a single problem instance. Points lying on the diagonal indicate that either the fixed value of $\zeta$ is the instance-specific optimal value or that it has a similar performance. Notably, for $\zeta=0.6$, the majority of points cluster tightly around the diagonal across all methods, suggesting that this choice often yields near-optimal performance. In contrast, for $\zeta=0.4$ and $\zeta=0.8$, many points fall noticeably below the diagonal, indicating that fixing $\zeta$ to these values more frequently results in suboptimal performance.

\section*{Supplementary Note 9: Extending the connection between $\zeta$ and the embedding validity to larger problems}
\begin{figure}[htb]
    \centering
    \includegraphics[width=1.0\linewidth]{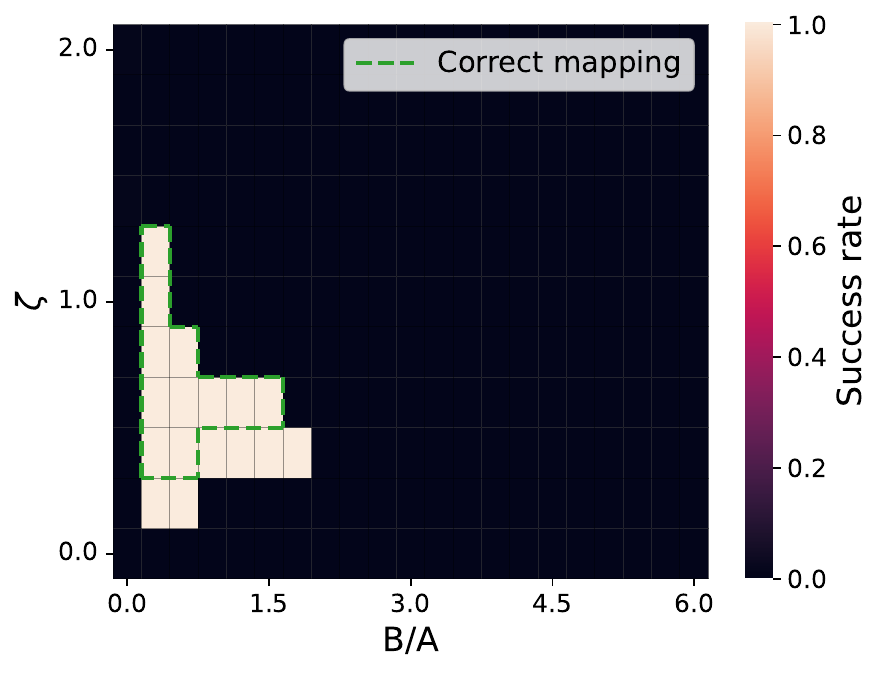}
    \caption{{\bf Correlation between success rate and correct mapping.} 
    Success rate of the g05\_10.0 graph when rescaling the external fields with a constant $\zeta$, as a function of  $\zeta$ and $B/A$ when slow annealing is performed ($d\beta/dt = 0.001$). No dynamic rescaling factor is applied. Other parameters are set as follows: $\alpha=-1$, $\gamma=0.001$, $n_\text{init}=100$.}
    \label{fig: max3cut: CIM SR g05u10d0}
\end{figure}

\begin{figure}[htb]
    \centering
    \includegraphics[width=1.0\linewidth]{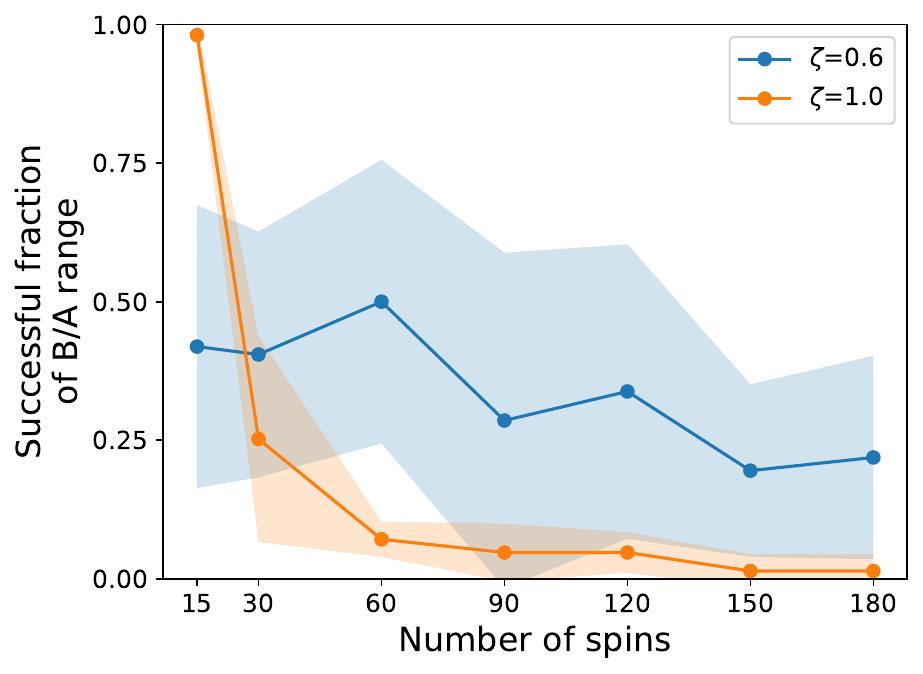}
    \caption{{\bf Fraction of valid $\boldsymbol{B/A}$ values.} Fraction of the tested $\frac{B}{A}$ values (defined in \cref{tab:OptimalParameters}) for which the success rate exceeds an arbitrary threshold of 10\% as a function of the number of spins. Each dot represents the mean value across 10 Max-3-Cut problems. Shaded areas represent the standard deviation. The underlying simulations were performed for the original external fields method, $\alpha=-1$, and $\frac{d\beta}{dt}=0.001$.
    }
    \label{fig: max3cut: viableBrange wrt N}
\end{figure}

Due to limited computational resources, it is not feasible to generate mapping-correctness plots like Fig.~6 from the main text for systems larger than 30 spins. However, as the IM is only expected to perform well when the mapping is correct, regions of high success rate can serve as a proxy for identifying regions of valid mapping. As illustrated in \cref{fig: max3cut: CIM SR g05u10d0}, the success rate under slow annealing conditions strongly correlates with the correct-mapping region identified in Fig.~6 of the main text. Notably, using $\zeta=0.4$ or $\zeta=0.6$ allows for a broader range of $B/A$ values that yield high success rates. To quantify this trend, \cref{fig: max3cut: viableBrange wrt N} shows the fraction of tested $B/A$ values (defined in \cref{tab:OptimalParameters}) that result in success rates larger than 10\% as a function of system size. The figure reveals that this fraction shrinks with increasing problem size for $\zeta=1$, while such a strong decrease is not observed for $\zeta=0.6$. Even though these figures are made using the original external fields, the conclusions also hold for the other methods, as the mapping correctness is independent of the method of implementing the external field. Moreover, the conclusions also hold when varying the arbitrary threshold of $10\%$ success rate.

\section*{Supplementary Note 10: optimal working point for the spin sign method.}
\begin{figure*}[htb]
    \centering
    \includegraphics[width=1.0\linewidth]{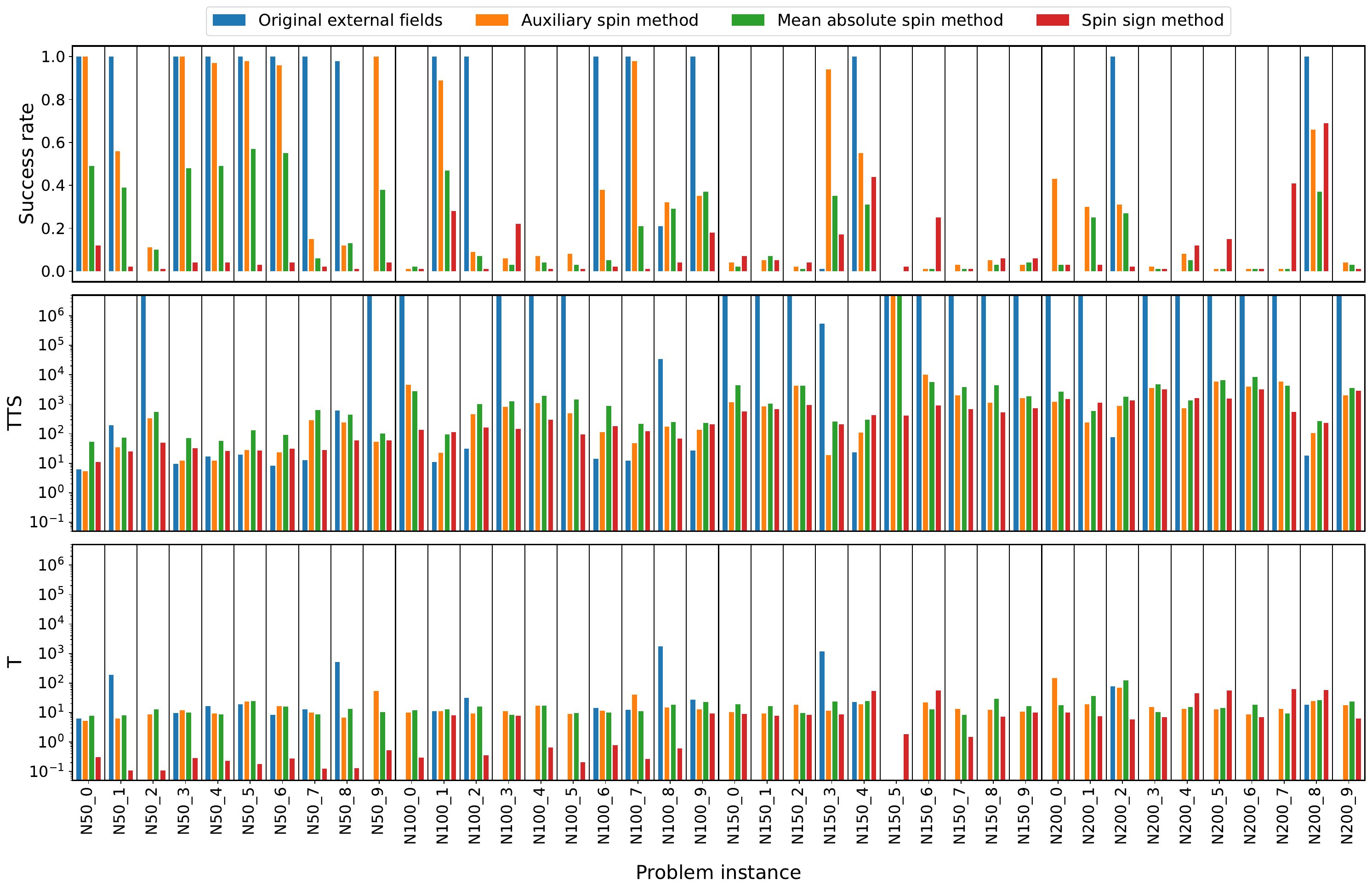}
    \caption{{\bf Performance for SK problems.}
    For every SK problem instance and for every implementation method for the external fields, the values of $\alpha$ and $v_\beta$ are determined that minimize TTS. For those values of the hyperparameters, the success rate, time-to-solution (TTS), and average runtime $T$ of successful runs is shown. TTS bars that reach the upper edge denote TTS=$\infty$.
    }
    \label{fig: SK: min TTS: SR TTS T}
\end{figure*}

\begin{figure*}[htb]
    \centering
    \includegraphics[width=1.0\linewidth]{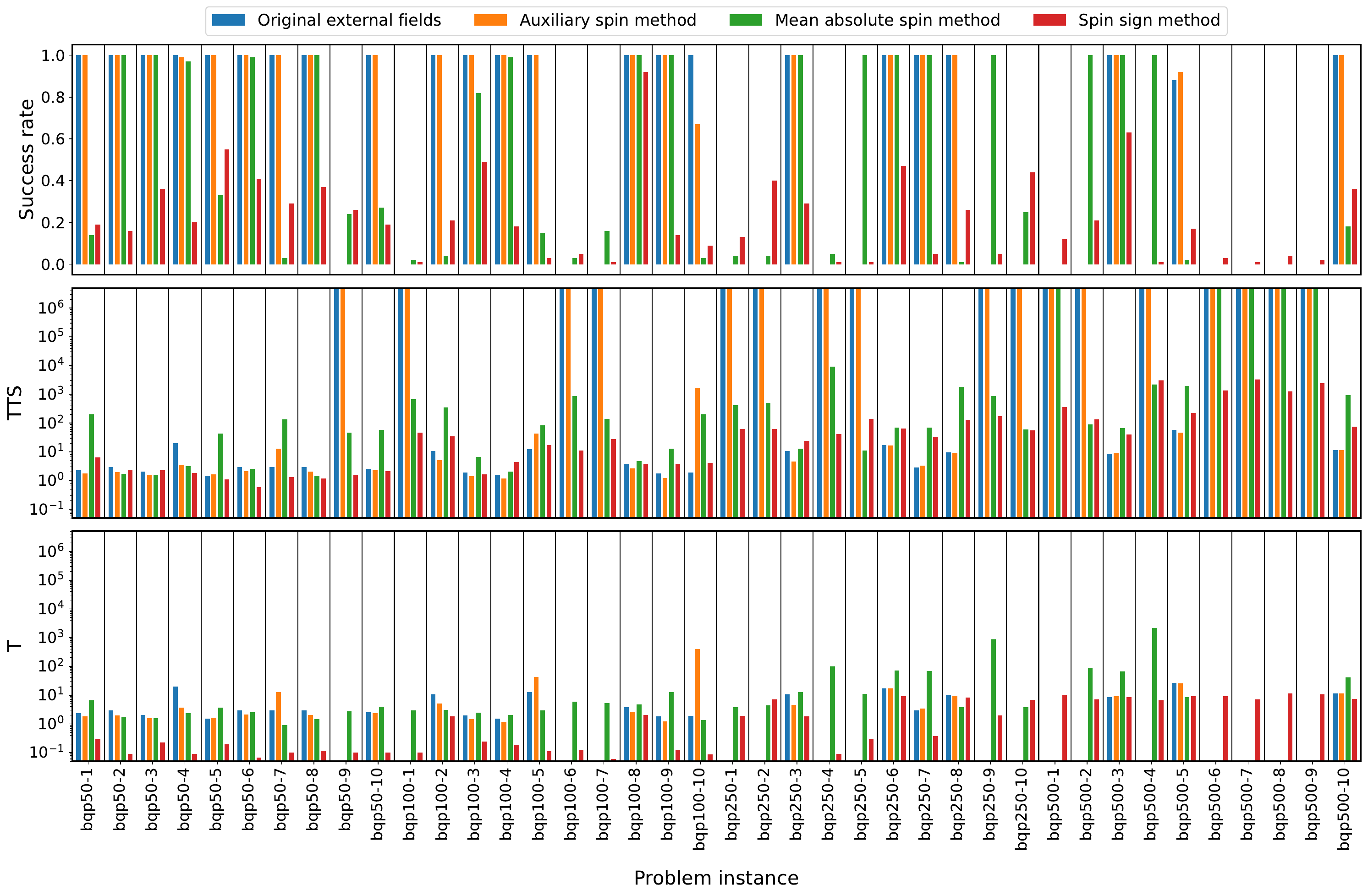}
    \caption{{\bf Performance for Beasley problems.}
    For every Beasley problem instance and for every implementation method for the external fields, the values of $\alpha$ and $v_\beta$ are determined that minimize TTS.For those values of the hyperparameters, the success rate, time-to-solution (TTS), and average runtime $T$ of successful runs is shown. TTS bars that reach the upper edge denote TTS=$\infty$.}
    \label{fig: beasley: min TTS: SR TTS T}
\end{figure*}

\begin{figure*}[htb]
    \centering
    \includegraphics[width=1.0\linewidth]{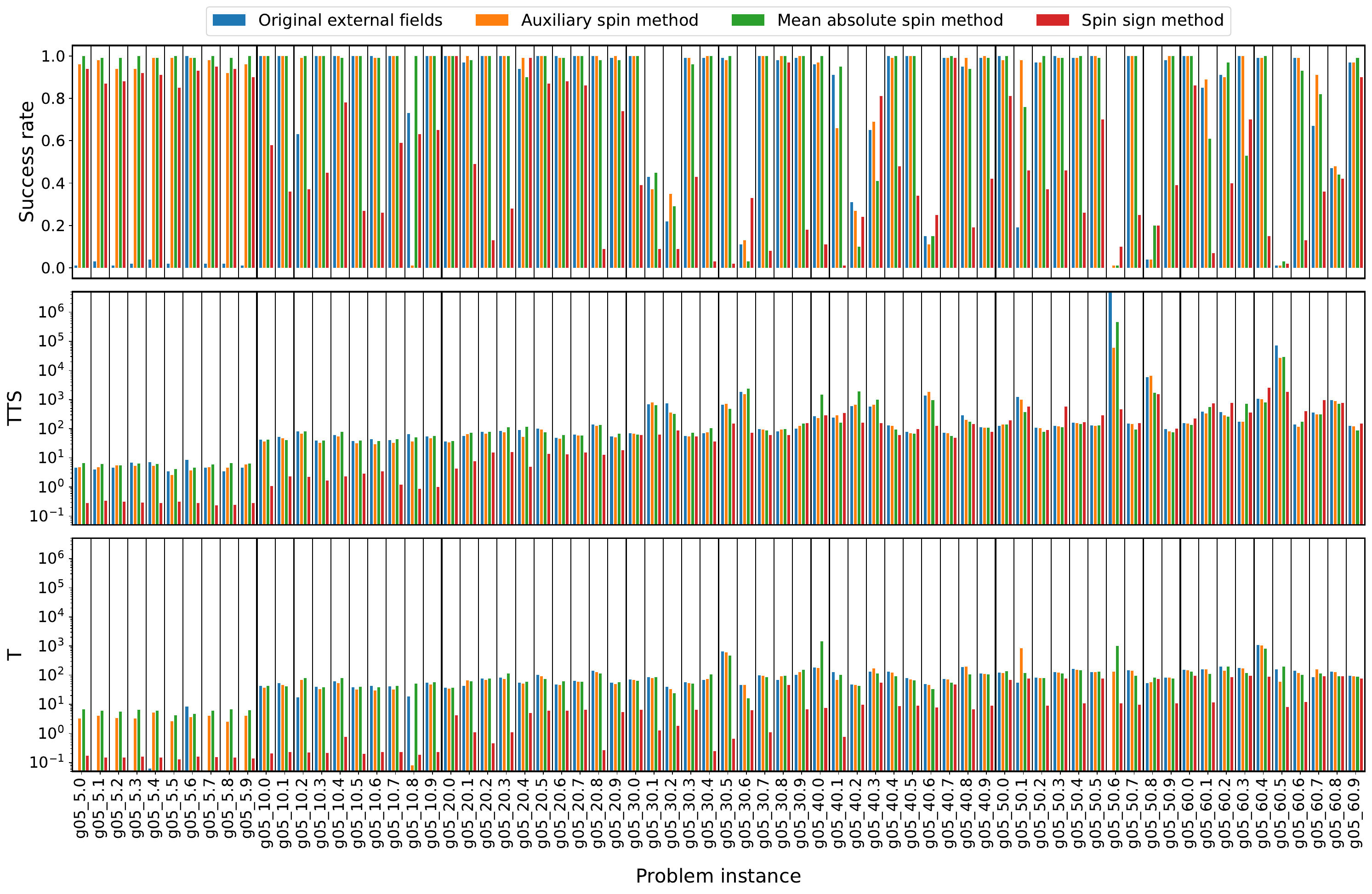}
    \caption{{\bf Performance for Max-3-Cut problems.}
    For every Max-3-Cut problem instance and for every implementation method for the external fields, the values of $\alpha$ and $v_\beta$ are determined that minimize TTS.For those values of the hyperparameters, the success rate, time-to-solution (TTS), and average runtime $T$ of successful runs is shown. TTS bars that reach the upper edge denote TTS=$\infty$.}
    \label{fig: max3cut: min TTS: SR TTS T}
\end{figure*}

\cref{fig: SK: min TTS: SR TTS T,fig: beasley: min TTS: SR TTS T,fig: max3cut: min TTS: SR TTS T} present the success rate, TTS, and average runtime $T$ of successful runs for all problem instances. \cref{fig: SK: min TTS: SR TTS T} corresponds to the SK instances, where the number following ``N" in the label indicates the number of spins. \cref{fig: beasley: min TTS: SR TTS T,fig: max3cut: min TTS: SR TTS T} report results for the Beasley and Max-3-Cut problems, respectively, using the BiqMac naming convention.

Across these figures, the spin-sign method often exhibits substantially lower success rates than the other three methods. However, its significantly shorter average runtime frequently results in superior TTS performance. This is because, when optimizing for TTS, the hyperparameter optimization tends to favor scenarios in which the spin-sign method rapidly converges, albeit with lower success probabilities, leading to a lower overall TTS. 
This type of operation is sometimes referred to as \textit{fail fast mode} \cite{noori2025statistical}.
Note that it is more often observed for smaller problem sizes. For larger problems, $T$ tends to be more similar between all methods.

\begin{figure*}[htb]
    \centering
    \includegraphics[width=1.0\linewidth]{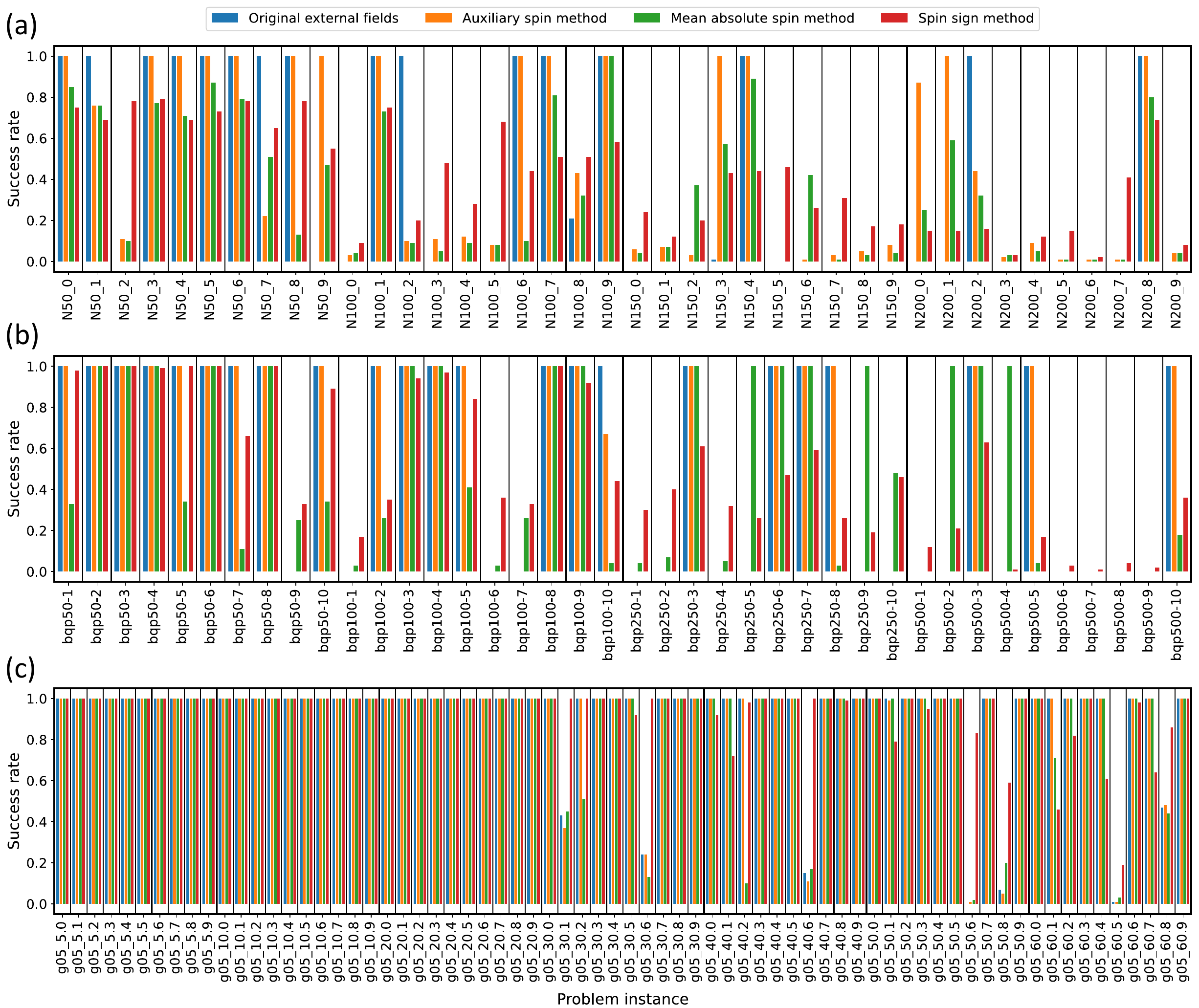}
    \caption{{\bf Performance for all problems when prioritizing success rate.}
    Results with hyperparameters chosen to maximize success rate. This contrasts with the main text, where hyperparameters are chosen to minimize TTS. Panels (a) and (b) show $\max_{\alpha, v_\beta} \text{SR}$ for SK and Beasley problems, respectively, with $\zeta = 1$. Panel (c) shows $\max_{\alpha, v_\beta, \frac{B}{A}} \text{SR}$ for Max-3-Cut problems, with $\zeta = 0.6$.}
    \label{fig: all COPs: max SR}
\end{figure*}

It is also important to note that the spin-sign method does not universally underperform in terms of success rate. While its success rates can be lower when performing a hyperparameter optimization that minimizes TTS, this is a consequence of favoring fast-converging configurations rather than a fundamental limitation of the method. In fact, several instances are solvable only with this method, while the others fail. Furthermore, as shown in \cref{fig: all COPs: max SR}, when the hyperparameters are optimized to maximize success rate instead of TTS, the spin-sign method operates in a regime with consistently higher success rates than before, often comparable to or higher than those of the other methods.

\end{document}